\documentclass[10pt,journal,compsoc]{IEEEtran}
\IEEEoverridecommandlockouts

\usepackage{fancyref}
\usepackage{dsfont}
\usepackage[utf8]{inputenc}
\usepackage[T1]{fontenc}
\usepackage{xspace}
\usepackage{amssymb}
\usepackage{amsthm}
\usepackage{xcolor}
\usepackage{lscape}
\usepackage{url}
\usepackage{multirow}
\usepackage{array}
\usepackage{paralist}
\usepackage{makecell}
\usepackage{adjustbox}
\usepackage[framemethod=tikz]{mdframed}
\usepackage[inline]{enumitem}
\usepackage{booktabs}
\usepackage{caption} 
\usepackage{icomma}
\usepackage{fp}
\usepackage[autolanguage]{numprint}

\usepackage{hhline}

\usepackage{colortbl}
\definecolor{lavender}{rgb}{0.9, 0.9, 0.98}
\newcommand*\np[2][z]{
\ifx z#1%
$\numprint{#2}$%
\else%
$\numprint[#1]{#2}$%
\fi\xspace
}

\newtheorem{definition}{Definition}

\newcommand{\revision}[1]{\textcolor{black}{#1}}

\usepackage[scaled=0.85]{beramono}
\usepackage{listings}
\usepackage{multicol}
\usepackage{amsmath}
\usepackage{microtype}
\usepackage[hidelinks, bookmarks=false]{hyperref}

\usepackage{algorithmic}
\usepackage{float}
\floatstyle{plaintop}
\restylefloat{table}
\usepackage{stfloats}
\usepackage{tabularx}
\usepackage{lipsum}
\usepackage{graphicx}
\ifCLASSOPTIONcompsoc
    \usepackage[caption=false, font=normalsize, labelfont=sf, textfont=sf]{subfig}
\else
\usepackage[caption=false, font=footnotesize]{subfig}
\fi

\usepackage{epstopdf} 
\usepackage{wrapfig}
\usepackage{pbox}
\usepackage{soul}
\usetikzlibrary{arrows.meta}
\usetikzlibrary{arrows}
\usetikzlibrary{shapes}
\usetikzlibrary{patterns}
\usetikzlibrary{positioning}
\usetikzlibrary{calc}
\usepackage{ragged2e}

\usepackage{colortbl}
\definecolor{pblue}{rgb}{0.13,0.13,1}
\definecolor{pgreen}{rgb}{0,0.5,0}
\definecolor{pred}{rgb}{0.9,0,0}
\definecolor{pgrey}{rgb}{0.46,0.45,0.48}
\usepackage{pifont} 

\usepackage{tikz}

\usepackage[english]{babel}
\addto\extrasenglish{
  
}
\addto\extrasenglish{
  
}
\addto\extrasenglish{
  
}

\usepackage{scalerel}
\usepackage{tikz}
\usetikzlibrary{svg.path}

\definecolor{orcidlogocol}{HTML}{A6CE39}
\tikzset{
  orcidlogo/.pic={
    \fill[orcidlogocol] svg{M256,128c0,70.7-57.3,128-128,128C57.3,256,0,198.7,0,128C0,57.3,57.3,0,128,0C198.7,0,256,57.3,256,128z};
    \fill[white] svg{M86.3,186.2H70.9V79.1h15.4v48.4V186.2z}
                 svg{M108.9,79.1h41.6c39.6,0,57,28.3,57,53.6c0,27.5-21.5,53.6-56.8,53.6h-41.8V79.1z M124.3,172.4h24.5c34.9,0,42.9-26.5,42.9-39.7c0-21.5-13.7-39.7-43.7-39.7h-23.7V172.4z}
                 svg{M88.7,56.8c0,5.5-4.5,10.1-10.1,10.1c-5.6,0-10.1-4.6-10.1-10.1c0-5.6,4.5-10.1,10.1-10.1C84.2,46.7,88.7,51.3,88.7,56.8z};
  }
}

\newcommand\orcidicon[1]{\href{https://orcid.org/#1}{\mbox{\scalerel*{
\begin{tikzpicture}[yscale=-1,transform shape]
\pic{orcidlogo};
\end{tikzpicture}
}{|}}}}

\lstdefinelanguage{Pom}{
  morekeywords={<dependency>,</dependency>,<groupId>,</groupId>,<artifactId>,</artifactId>,<version>,</version>,<scope>,</scope>},
  otherkeywords={<dependency>,</dependency>,<groupId>,</groupId>,<artifactId>,</artifactId>,<version>,</version>,<scope>,</scope>}
}

\lstdefinelanguage{json}{
    basicstyle=\footnotesize\ttfamily,
    numbers=left,
    numberstyle=\scriptsize,
    stepnumber=1,
    numbersep=8pt,
    showstringspaces=false,
    breaklines=true,
    frame=lines,
    backgroundcolor=\color{white},
    stringstyle=\color{blue},
    commentstyle=\color{gray},
    keywordstyle=\color{purple},
    morekeywords={true,false,null}
}

\usetikzlibrary{calc}
\usetikzlibrary{decorations.pathmorphing}

\usepackage{booktabs}
\def\BibTeX{{\rm B\kern-.05em{\sc i\kern-.025em b}\kern-.08emT\kern-.1667em\lower.7ex\hbox{E}\kern-.125emX}}

\mdfdefinestyle{mpdframe}{
    frametitlebackgroundcolor   =black!15,
    backgroundcolor             =black!3,
    frametitlerule              =true,
    roundcorner                 =3pt,
    middlelinewidth             =0.8pt,
    innermargin                 =0.2cm,
    outermargin                 =0.2cm,
    innerleftmargin             =0.2cm,
    innerrightmargin            =0.2cm,
    innertopmargin              =0.2cm,
    innerbottommargin           =0.2cm,
    nobreak=true
}

\input{utils/er.tex}  
\input{utils/abbrev}
\definecolor{pblue}{rgb}{0.13,0.13,1}
\definecolor{pgreen}{rgb}{0,0.5,0}
\definecolor{pred}{rgb}{0.9,0,0}
\definecolor{pgrey}{rgb}{0.46,0.45,0.48}
\definecolor{ashgrey}{rgb}{0.7, 0.75, 0.71}
\usepackage{listings}
  \lstdefinelanguage{diff}{
    basicstyle=\ttfamily\small,
    morecomment=[f][\color{pgrey}]{@@},
    morecomment=[f][\color{pred}]{+\ },
    morecomment=[f][\color{pgreen}]{-\ },
  }

\makeatletter
\newenvironment{btHighlight}[1][]
{\begingroup\tikzset{bt@Highlight@par/.style={#1}}\begin{lrbox}{\@tempboxa}}
{\end{lrbox}\bt@HL@box[bt@Highlight@par]{\@tempboxa}\endgroup}
\newcommand\btHL[1][]{%
  \begin{btHighlight}[#1]\bgroup\aftergroup\bt@HL@endenv%
}
\def\bt@HL@endenv{%
  \end{btHighlight}%
  \egroup
}
\newcommand{\bt@HL@box}[2][]{%
  \tikz[#1]{%
    \pgfpathrectangle{\pgfpoint{1pt}{0pt}}{\pgfpoint{\wd #2}{\ht #2}}%
    \pgfusepath{use as bounding box}%
    \node[anchor=base west, fill=orange!30,outer sep=0pt,inner xsep=1pt, inner ysep=0pt, rounded corners=0pt, minimum height=\ht\strutbox+1pt,#1]{\raisebox{1pt}{\strut}\strut\usebox{#2}};
  }%
}
\lstdefinestyle{Java}{
    language={Java}, 
    moredelim=**[is][{\btHL[fill=red!17,thin]}]{`}{`},
    moredelim=**[is][{\btHL[fill=green!17,thin]}]{@}{@},
    moredelim=**[is][{\btHL[fill=yellow!17,thin]}]{~}{~},
}

\newboolean{showcomments}
\setboolean{showcomments}{true}
\ifthenelse{\boolean{showcomments}}
{
}

\newcommand{\ShowAbsoluteNumber}[1]{%
\ifnum #1<10%
{\hspace*{0pt}#1}%
\else%
\ifnum #1<100%
{\hspace*{0pt}#1}%
\else%
\ifnum #1<1000%
{\hspace*{0pt}#1}%
\else%
{\numprint{#1}}%
\fi%
\fi%
\fi%
}

\newcommand{\ShowPercentage}[2]{%
\FPeval\percentage{round(#1/#2*100,0)}%
\FPeval\percentageOneDecimal{round(#1/#2*100,1)}%
\ifnum \percentage=0%
{\np[\%]{\FPprint{percentageOneDecimal}}}%
\else%
\ifnum \percentage<10%
{\np[\%]{\FPprint{percentageOneDecimal}}}%
\else%
{\np[\%]{\FPprint{percentageOneDecimal}}}%
\fi%
\fi%
\xspace
}
\newlength\BARSIZE  
\newcommand{\inlinechart}[2]{%
\FPeval{\BLACKBARSIZE}{#1/#2}\textcolor{black!80}{\rule{\BLACKBARSIZE\BARSIZE}{1.6ex}}%
\FPeval{\BLACKBARSIZE}{1 - (#1/#2)}\textcolor{black!10}{\rule{\BLACKBARSIZE\BARSIZE}{1.6ex}}%
}

\newcommand*\ChartSmall[3][v]{%
\ifx q#1%
    \np{#2}/\np{#3}(\ShowPercentage{#2}{#3})\else%
\ifx p#1%
    \np{#2}(\ShowPercentage{#2}{#3})\else%
\ifx c#1%
    \inlinechart{#2}{#3}%
\else%
    \np{#2}%
    \ifx r#1%
        /\np{#3}%
    \fi%
    \hspace*{0.5ex}(\ShowPercentage{#2}{#3}) %
    \inlinechart{#2}{#3}%
    \xspace
\fi\fi\fi%
}

\newcommand{\Ratio}[2]{
  \FPeval\percentage{round(#2/#1,1)}
  \FPeval\percentageOneDecimal{round(#2/#1,1)}
  \np[\times]{\FPprint{percentageOneDecimal}}
}

\newcolumntype{?}{!{\vrule width 1.2pt}}
\usepackage{colortbl}

\usepackage{balance}
\usepackage{flushend}

\usepackage{algorithm}

\begin{document}



\def\ratioClassesInProjectWrtClassesInDependenciesOriginal{\textcolor{black}{\Ratio{15594}{135343}}}
\def\ratioClassesInProjectWrtClassesInDependenciesBeforeDeptrim{\textcolor{black}{\Ratio{15594}{128603}}}
\def\ratioClassesInProjectWrtClassesInDependenciesAfterDeptrim{\textcolor{black}{\Ratio{15594}{67722}}}
\def\nbNBCD{\textcolor{black}{\np{396}}\@\xspace}
\def\nbClassesInJacopProject{\textcolor{black}{\np{833}}\@\xspace}

\def\ratioDebloated{\textcolor{black}{\Ratio{15594}{129530}}}
\def\perOfClassesDebloated{\textcolor{black}{\ChartSmall[p]{5813}{135343}}\@\xspace}
\def\percNBCD{\textcolor{black}{\ChartSmall[p]{396}{467}}\@\xspace}
\def\percBloatedDependenciesRemoved{\textcolor{black}{\ChartSmall[p]{71}{467}}\@\xspace}
\def\perOfClassesRemovedWithDepclean{\textcolor{black}{\ShowPercentage{5813}{135343}}\@\xspace}
\def\nbOfDependenciesDebloated{\textcolor{black}{71}\@\xspace}

\def\ratioTstThatPass{\textcolor{black}{\ChartSmall[p]{14}{30}}\@\xspace}
\def\nbTstThatPass{\textcolor{black}{\np{14}}\@\xspace}
\def\percTstThatPass{\textcolor{black}{\ShowPercentage{14}{30}}\@\xspace}
\def\nbTstThatFail{\textcolor{black}{\np{16}}\@\xspace}

\def\ratioPstThatPass{\textcolor{black}{\ChartSmall[p]{16}{30}}\@\xspace}
\def\nbPstThatPass{\textcolor{black}{\np{16}}\@\xspace}
\def\percPstThatPass{\textcolor{black}{\ShowPercentage{16}{30}}\@\xspace}
\def\nbPstThatFail{\textcolor{black}{\np{16}}\@\xspace}

\def\nbSpecializedDependencies{\textcolor{black}{\np{343}}\@\xspace}
\def\ratioOfSpecializedDependencies{\textcolor{black}{\ChartSmall[p]{343}{396}}\@\xspace}

\def\nbTotalClassesRemoved{\textcolor{black}{\np{51631}}\@\xspace}
\def\nbTotalSizeReduction{\textcolor{black}{\np[KB]{152231}}\@\xspace}

\def\ratioClassesRemoved{\textcolor{black}{\ChartSmall[p]{51631}{129530}}\@\xspace}
\def\ratioSizeReduction{\textcolor{black}{\ChartSmall[p]{152231}{268985}}\@\xspace}

\def\percOfClassesRemoved{\textcolor{black}{\ShowPercentage{51631}{129530}}\@\xspace}
\def\percOfSizeReduction{\textcolor{black}{\ShowPercentage{152231}{268985}}\@\xspace}

\def\ratioDeptrim{\textcolor{black}{\Ratio{15594}{77899}}}

\def\ratioPstCompilationErrors{\textcolor{black}{\ChartSmall[p]{12}{364}}\@\xspace}
\def\ratioPstTestFailures{\textcolor{black}{\ChartSmall[p]{9}{364}}\@\xspace}

\def\nbPstWithCompilationErrors{\textcolor{black}{\np{11}}\@\xspace}
\def\nbPstWithTestFailures{\textcolor{black}{\np{9}}\@\xspace}

\def\nbProjectsWithCompilationErrors{\textcolor{black}{\np{11}}\@\xspace}
\def\nbProjectsWithTestFailures{\textcolor{black}{\np{9}}\@\xspace}

\def\ratioPstCompilationPass{\textcolor{black}{\ChartSmall[p]{352}{364}}\@\xspace}
\def\ratioPstTestPass{\textcolor{black}{\ChartSmall[p]{387}{396}}\@\xspace}

\def\nbTotalTestsExecuted{\textcolor{black}{\np{27844}}\@\xspace}
\def\nbTotalTestsExecutedThatFail{\textcolor{black}{\np{130}}\@\xspace}

\def\PercentageOfTestsExecutedThatFail{\textcolor{black}{\ShowPercentage{130}{11912}}\@\xspace}

\def\nbTUD{\textcolor{black}{\np{32}}\@\xspace}
\def\PercentageOfTUD{\textcolor{black}{\ShowPercentage{32}{396}}\@\xspace}

\title{Automatic Specialization of Third-Party\\ Java Dependencies}




\author{
\IEEEauthorblockN{
	\normalsize
	C\'esar Soto-Valero$^{~\orcidicon{0000-0002-1996-6134}}$,
	Deepika Tiwari$^{~\orcidicon{0000-0003-0293-2592}}$,
     Tim Toady$^{~\orcidicon{0000-0002-0209-2805}}$,
	   and Benoit Baudry    $^{~\orcidicon{0000-0002-4015-4640}}$}

\IEEEauthorblockA{
\textit{KTH Royal Institute of Technology, Stockholm, Sweden}\\
Email: \{cesarsv, deepikat, baudry\}@kth.se; toady@eecs.kth.se\\
}
}

\IEEEtitleabstractindextext{
\begin{abstract}
Large-scale code reuse significantly reduces both development costs and time. However, the massive share of third-party code in software projects poses new challenges, especially in terms of maintenance and security.
In this paper, we propose a novel technique to specialize dependencies of Java projects, based on their actual usage.
Given a project and its dependencies, we systematically identify the subset of each dependency that is necessary to build the project, and we remove the rest. As a result of this process, we package each specialized dependency in a \jar file. 
Then, we generate specialized dependency trees where the original dependencies are replaced by the specialized versions. This allows building the project with significantly less third-party code than the original.
As a result, the specialized dependencies become a first-class concept in the software supply chain, rather than a transient artifact in an optimizing compiler toolchain.
We implement our technique in a tool called \deptrim, which we evaluate with \np{30} notable open-source Java projects.
\deptrim specializes a total of  \ratioOfSpecializedDependencies dependencies across these projects, and successfully rebuilds each project with a specialized dependency tree.
Moreover, through this specialization, \deptrim removes a total of \np{57444} (\np[\%]{42.2}) classes from the dependencies, reducing the ratio of dependency classes to project classes from \ratioClassesInProjectWrtClassesInDependenciesOriginal in the original projects to \ratioDeptrim after specialization. These novel results indicate that dependency specialization significantly reduces the share of third-party code in Java projects.\looseness=-1 
\end{abstract}

}

\maketitle
\thispagestyle{plain}
\pagestyle{plain}

\IEEEdisplaynontitleabstractindextext
\IEEEpeerreviewmaketitle

\let\url\nolinkurl

\section{Introduction}\label{sec:introduction}


\IEEEPARstart{S}{oftware} projects are developed by assembling new features and components provided by reusable third-party libraries. 
Software reuse at large is a known best practice in software engineering \cite{krueger1992software}. 
Its adoption has rocketed in the last decade, thanks to the rapid growth of repositories of reusable packages, along with the development of mature package managers \cite{cox2019surviving}. 
These package managers let developers declare a list of third-party libraries that they want to reuse in their projects. 
The libraries declared by developers form the set of direct dependencies of the project. Then, at build time, the package manager fetches the code of these libraries, as well as the code of transitive dependencies, declared by the direct dependencies.
This forms a dependency tree that the build system bundles with the project code into a package that can be released and deployed. 

The large-scale adoption of software reuse \cite{gustavsson2020managing} is beneficial for software companies as it reduces their delivery times and costs \cite{harutyunyan2020managing}. Meanwhile, reuse today has reached a point where most of the code in a released application actually originates from third-party dependencies \cite{ombredanne2020free}. 
This massive presence of third-party code in application binaries has turned software reuse into a double-edged sword
\cite{gkortzis2021software}.
Recent studies have highlighted the new challenges that third-party dependencies pose for maintenance \cite{DecanMC18,JafariCAST22}, performance \cite{heo2018effective}, code quality \cite{WangWWWLCYXZ22}, and security \cite{DannPHPB22,PashchenkoPPSM22}.

Several techniques have emerged to address the challenges of dependency management. 
The first type of approach consists of supporting developers in maintaining a correct and secure dependency tree. 
Software composition analysis \cite{imtiaz2021comparative} and software bots \cite{Mahmoud21} suggest dependency updates and warn about potential vulnerabilities among dependencies. 
Integrity-checking tools aim at preventing packaging a dependency with code that may have been tempered with. For example, the Go community maintains a global  database for authenticating module content \cite{cox2022go} 
and sigstore facilitates the procedure of signing third-party libraries~\cite{newman2022sigstore}.
A second type of approach to maintain healthy dependency trees consists in reducing it, removing the dependencies that are completely unused. Examples of such techniques include package debloating for Linux applications \cite{pashakhanloo2022pacjam} dependency debloating or shading for Java applications \cite{ponta2021used}, or tree shaking for JavaScript applications \cite{LatendresseMCS22}.

In this paper, we aim at advancing the state-of-the-art of dependency tree reduction with  a novel technique that specializes dependency trees to the needs of an application.
Mishra and Polychronakis first  introduced the concept of API specialization \cite{mishra2018shredder} to reduce the attack surface of third-party libraries. In this work, we advance the state-of-the-art on library specialization with a combination of static and dynamic code analysis to generate (i) a specialized version of an application's dependency tree and (ii) specialized versions of each third-party library that is partially used by an application, which can be deployed in a library registry. We implement our new technique in \deptrim, a tool that automatically specializes third-party libraries in the dependency tree of Java applications.

\deptrim analyses the bytecode of a Java project, as well as all its direct and transitive third-party dependencies. First, it removes the dependencies that are completely bloated, and identifies the non-bloated ones. \revision{Next, for each non-bloated dependency, \deptrim builds a static call graph through all non-bloated dependencies, to identify the classes for which at least one member is reachable from the project. \deptrim then removes the unused classes and produces one specialized jar for each  dependency. 
Finally, \deptrim modifies the dependency tree of the project,  replacing the original dependencies with the specialized versions in the build file.} 
The output of \deptrim is a specialized dependency tree of the project, with the maximum number of specialized dependencies such that the project builds correctly, \revision{\ie, the project correctly compiles and all its tests pass, providing evidence that the expected behavior of the project, as specified within the test suite, is preserved. 
\deptrim simplifies the reuse of specialized dependencies by generating reusable \jar files, which can be readily deployed to external repositories and can be documented and versioned as part of the project's software bill of material \cite{xia2023}.}

We demonstrate the capabilities of \deptrim by performing a study with \np{30} mature open-source Java projects that are configured to  build with \maven. 
\deptrim successfully analyzes \np{135343} classes across the \np{467} dependencies of the projects.
For \nbTstThatPass projects, it generates a dependency tree in which all \compile-scope dependencies are specialized. For the \nbPstThatPass other projects, \deptrim produces a dependency tree that includes all dependencies that can be specialized without breaking the build, while keeping the others intact. 
In total, \deptrim removes \ratioClassesRemoved unused classes from \nbSpecializedDependencies third-party dependencies. 
The specialized dependencies are deployed locally, as reusable \jar files. For each project, \deptrim produces a specialized version of the \pom file that replaces original dependencies with specialized ones, such that the project still correctly builds.


In summary, our contributions are as follows:
\begin{itemize} 
    \item A fully automated technique to specialize the dependency tree of Java projects at build time.
    \item A tool called \deptrim, which automatically builds \maven projects with the largest subset of specialized dependencies. 
    \item Novel observations about  the ratio of dependency classes compared to project classes collected on \np{30} mature open-source projects at three stages of the dependency tree: original, debloated, and specialized.
    \item Empirical evidence that \deptrim successfully specializes the dependency tree of \nbTstThatPass projects in its entirety, and \nbPstThatPass partially, reducing the number of third-party classes by (\np[\%]{42.2}). The project classes to dependency classes ratio is divided by two, from \np{8.7}$\times$ to \np{5.0}$\times$. 
\end{itemize}


\section{Background}\label{sec:background}

In this section, we introduce the existing techniques to reduce the amount of dependency code.
Then we present the opportunities for dependency specialization, for Java projects.

\subsection{Terminology}

\begin{figure}[t]
  \centering
  \subfloat[\maven dependency resolution (default)\label{fig:jacop-dt-original}]{
       \includegraphics[width=8.5cm]{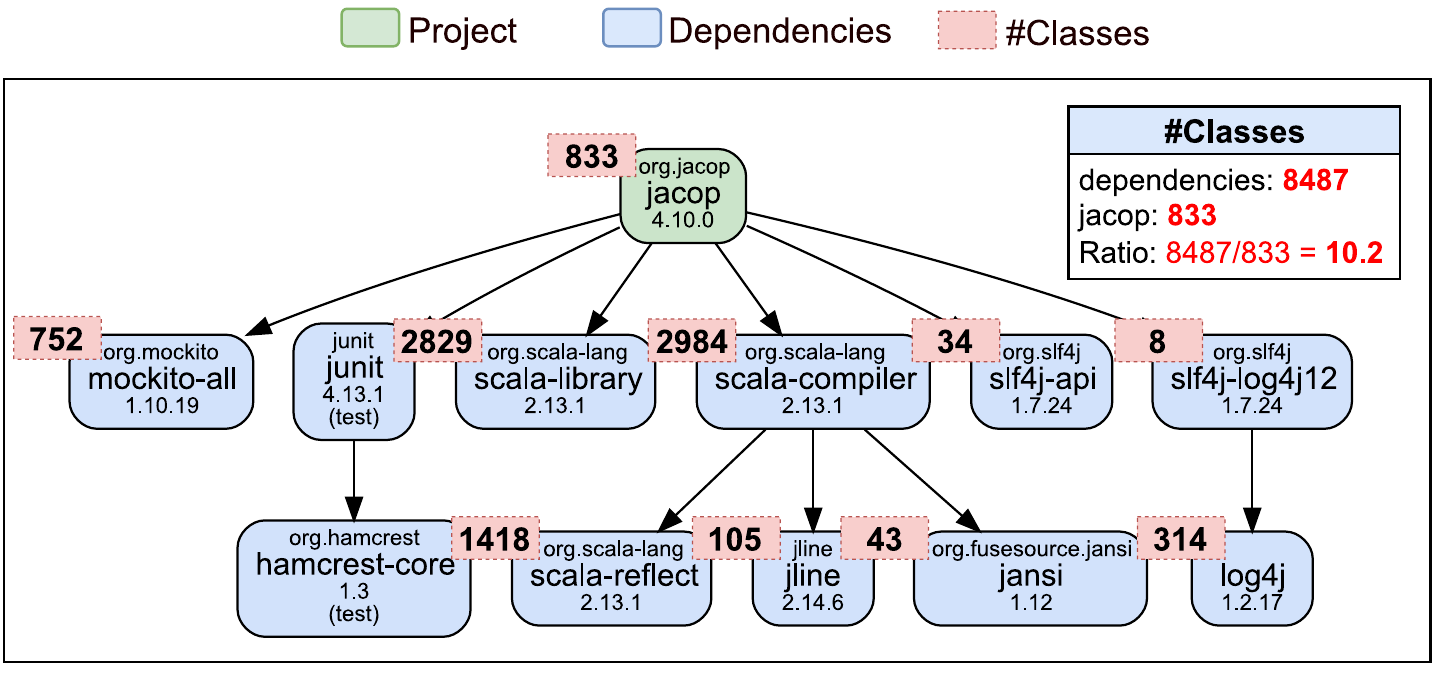}}
    \\
  \subfloat[\depclean dependency debloating (state-of-the-art)\label{fig:jacop-dt-debloated}]{
        \includegraphics[width=8.5cm]{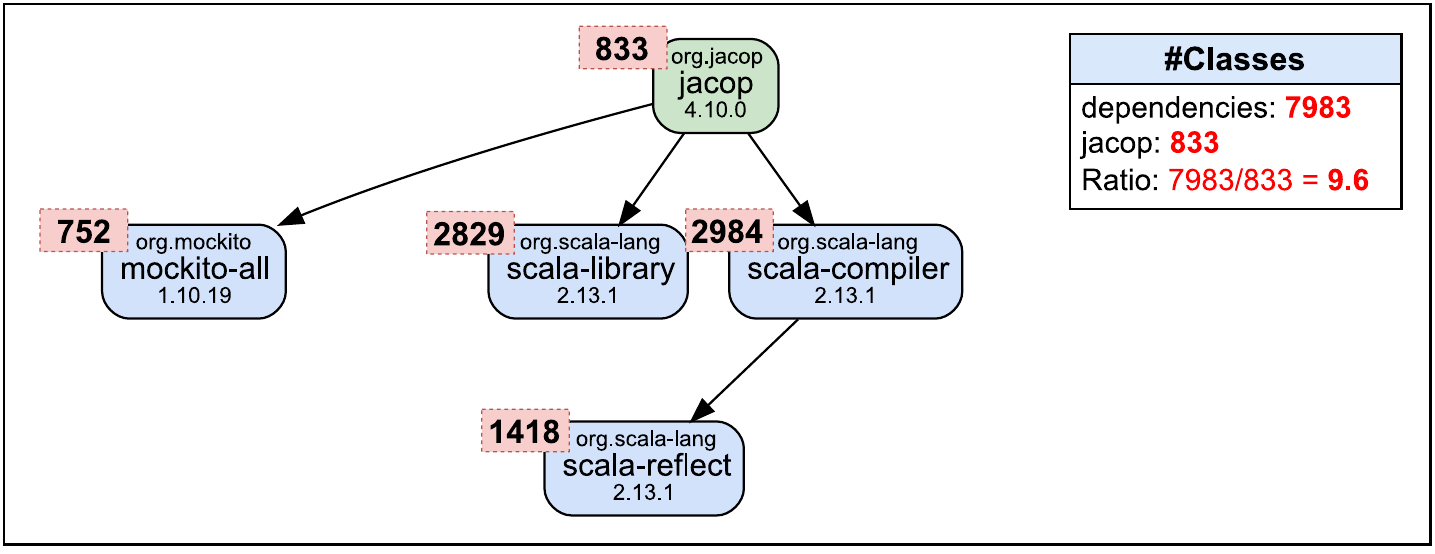}}
    \\
  \subfloat[Classes actually used in the dependencies (determined by static analysis)\label{fig:jacop-dt-specialized}]{
        \includegraphics[width=8.5cm]{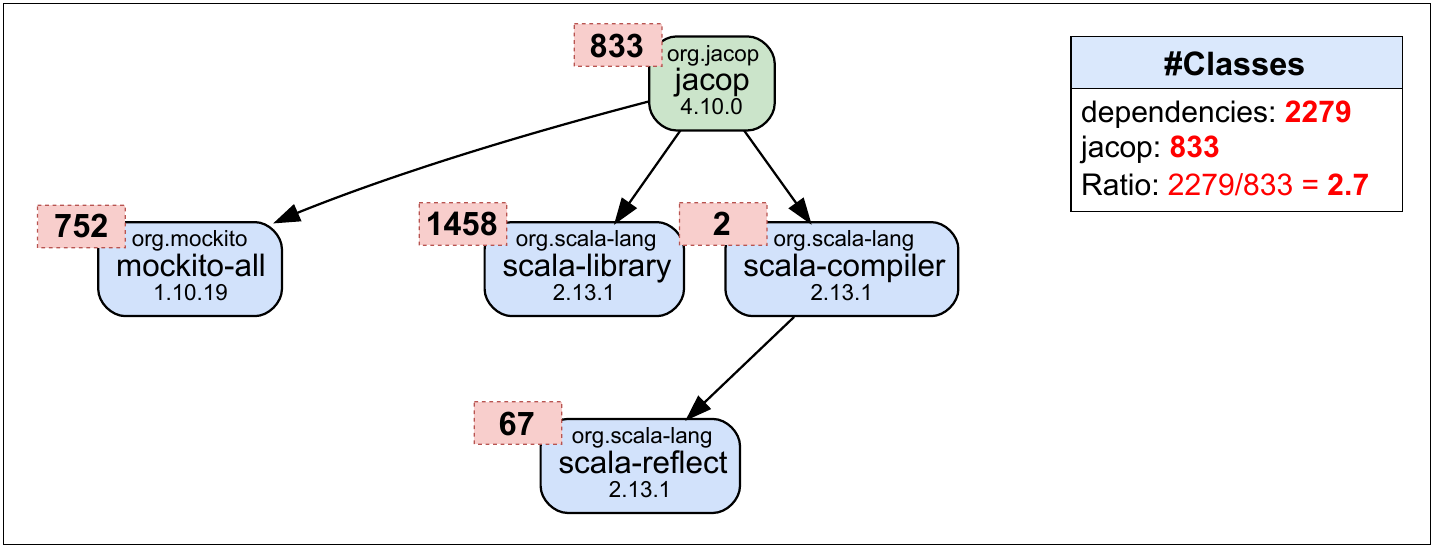}}
  \caption{Example of transformations to reduce the share of dependency code  in the project \jacop \texttt{v4.10.0}, and the impact of such transformations on the dependency classes to project classes ratio. Dependencies have \compile-scope by default if not specified. Note that \jacop reuses only a portion of its third-party dependencies.}
  \label{fig:jacop-dt}
  \vspace{-0.25cm}
\end{figure}

In this work, we consider a software project as a collection of Java source code files and configuration files organized to be built with \maven~\cite{ApacheMaven2023}.
\maven is a build automation tool for Java-based projects. 
It is primarily used for managing the dependencies of a project, testing it, and packaging it, as specified in the Project Object Model (POM) expressed in a file called \pom.
This file, located at the root of the project, includes additional information such as the project name and version.
We now define the key concepts about dependencies in the \maven ecosystem.

\begin{definition}
\textbf{Maven dependency: }\normalfont
A \maven dependency defines a relationship between a project and another compiled project.
Dependencies are compiled \jar files, uniquely identified with a triplet (\texttt{G:A:V}) where \texttt{G} is the \texttt{groupId}, \texttt{A} is the \texttt{artifactId}, and \texttt{V} is the \texttt{version}.
Dependencies are defined in the \pom within a scope, which determines the phase of the \maven build cycle at which the dependency is required.
\maven distinguishes \np{6} dependency scopes: \texttt{compile}, \texttt{runtime}, \texttt{test}, \texttt{provided}, \texttt{system}, and \texttt{import}.

For example, the constraint programming solver \jacop (\texttt{\href{https://github.com/radsz/jacop/commit/1a395e6add22caf79590fe9d1b2223bfb6ed0cd0}{6ed0cd0}}) is a \maven project. As illustrated in \autoref{fig:jacop-dt-original}, \texttt{scala-library} is one of its \np{11} dependencies. 
This is a \compile-scope dependency, which means that \jacop can use some functionalities of \texttt{scala-library} at compile time, and will include all the code of \texttt{scala-library} within the packaged binary of \jacop.
The testing framework \texttt{junit} is also declared as  a dependency of \jacop within the \texttt{test} scope, indicating that this dependency is required only for executing unit tests.

\end{definition}

\begin{definition}
\textbf{Dependency tree: }\normalfont
The dependency tree of a \maven project is a directed acyclic graph that includes all the direct dependencies declared  by developers in the project \pom, as well as all the transitive dependencies, \ie dependencies in the transitive closure of direct dependencies. For a \maven project, there exists a dependency resolution mechanism that fetches both direct and transitive dependency \texttt{JAR} files not present locally from external repositories such as Maven Central \cite{MavenDependencyMechanism2021}.
The project becomes the root node of the tree, while the edges represent dependency relationships between its direct and transitive dependencies.

For example, in \autoref{fig:jacop-dt-original}, \texttt{scala-compiler} is a direct dependency of \jacop because it is declared by developers in the \pom. It depends on \texttt{scala-reflect}, which makes \texttt{scala-reflect} a transitive dependency of \jacop. The \np{6} direct and \np{5} transitive dependencies of \jacop constitute its dependency tree.
\end{definition}

\begin{definition}
\textbf{Bloated dependency: }\normalfont
A dependency is said to be bloated if none of the elements in its API are used, directly or indirectly, by the project
~\cite{soto2021comprehensive}. This means that, although they are present in the dependency tree of software projects, bloated dependencies are entirely unused. Developers are therefore encouraged to remove them~\cite{soto2021longitudinal}.

For example, \autoref{fig:jacop-dt-debloated} presents the \compile-scope dependencies of \jacop after the removal of its bloated dependencies.  The dependencies \texttt{jline}, \texttt{jansi}, \texttt{sl4j-api}, \texttt{sl4j-log4j12}, and \texttt{log4j} are bloated and have been safely removed, as no member of their APIs is exercised by \jacop.
\end{definition}

\subsection{Example}

We now illustrate the \maven dependency resolution mechanism and the concept of bloated dependencies.
\autoref{fig:jacop-dt} shows the example of the transformations of the dependency tree of the project \jacop.
In \autoref{fig:jacop-dt-original}, we see the dependency tree of \jacop as generated by the \maven dependency resolution mechanism: it  fetches \jar files from external repositories  while omitting duplication, avoiding conflicts, and constructing a tree representation of the dependencies~\cite{wang2018dependency}.
\jacop has a total of \np{11} third-party dependencies: \np{6} are direct and \np{5} are transitive.
Direct dependencies are explicitly declared by the developers in the \pom file of \jacop, while transitive dependencies are resolved automatically via the \maven dependency resolution mechanism.
\maven uses the concept of scope to determine the visibility and lifecycle of a dependency, \ie, whether it should be included in the classpath of a certain build phase, as well as what the classpath of an artifact should be during the execution of a build phase.
For example, \jacop has \np{9} \compile-scope dependencies (the default) and \np{2} \texttt{test} scope dependencies. 
When \jacop is packaged for deployment as a \texttt{jar-with-dependencies}, its \jar file will include the bytecode of all its \np{9} \compile-scope dependencies. These \compile-scope dependencies include \np{8487} \class files, while the number of classes within \jacop, written and tested by its developers, is \np{833}. 
As observed, the number of classes contributed by third-party dependencies is one order of magnitude (i.e., \np{10.2}$\times$) more than the number of classes written by the \jacop developers.

When we run \depclean, a state-of-the-art \maven plugin that identifies and removes bloated dependencies~\cite{soto2021comprehensive,ponta2021used}, we find that \np{5} dependencies of \jacop are never used, and are therefore marked as bloated. 
\autoref{fig:jacop-dt-debloated} shows the dependency tree of \jacop after \texttt{test}-scope dependencies and bloated dependencies are removed. 
In this case, the number of nodes in the tree is reduced from \np{11} to \np{4}.
The reduction in the number of \compile-scope dependencies represents a removal of \ChartSmall[p]{504}{8587} third-party classes (\eg, removing \texttt{sl4j-api} leads to the removal of \np{34} classes).
For \jacop, the removal of bloated dependencies has a minimal impact on the reduction of third-party classes.
Consequently, while complete dependency debloating drastically reduces the number of dependencies in  \jacop, it only leads to a modest reduction in the ratio of dependency classes to project classes, from the original \np{10.2}$\times$ in \autoref{fig:jacop-dt-original} to \np{9.6}$\times$ in \autoref{fig:jacop-dt-debloated}.

To assess the opportunities of further reducing the number of dependency classes, we analyze the \jar of each non-debloated dependency of \jacop. 
We compute the static call graph of method calls between the classes in the \jar files. Based on this graph, we get the list of dependency classes that are reachable from the project at build time.
\autoref{fig:jacop-dt-specialized} shows the  number of reachable classes for each dependency of \jacop. 
Consider the direct dependency \texttt{scala-compiler}. Of its \np{2984} classes, only two are reachable from \jacop. 
This confirms that \texttt{scala-compiler} is not a bloated dependency for \jacop, and that it includes way more features than what \jacop actually needs.
This is evidence of the opportunity to specialize this dependency in the context of \jacop. 
Similar opportunities exist for \np{2} other non-bloated dependencies. 
In fact, we find that \ChartSmall[q]{5704}{7983} of the third-party classes in these dependencies can be removed, and \jacop can still build successfully.
After dependency specialization, the ratio of the number of dependency classes to \jacop's classes is \np{2.7}$\times$.
This is a drastic reduction from \np{9.6}$\times$ which was the ratio after debloating (\autoref{fig:jacop-dt-debloated}), and even more significant if we consider the original ratio of \np{10.2}$\times$  in \autoref{fig:jacop-dt-original}.  

The number of classes actually used in the dependencies is significantly lower than the original number of classes provided.
This observation motivates us to extend the state-of-the-art of Java dependency management with a novel technique to specialize non-bloated dependencies, by identifying and removing unnecessary classes through bytecode removal.
In the next section, we present our approach and provide details on \deptrim, a tool that automatically specializes the dependencies of \maven projects.

\section{Dependency Specialization With~\deptrim}\label{sec:tool}

This section presents \deptrim, an end-to-end tool for the automated specialization of third-party Java dependencies. We define the concept of dependency specialization, followed by an explanation of the key phases of \deptrim.

\subsection{Dependency Specialization}\label{sec:deptrim-definitions}
This work introduces the concept of specialized dependencies and specialized dependency trees. We define them below.

\begin{definition}
\textbf{Specialized dependency: }\normalfont
\revision{A dependency is said to be specialized with respect to a project if all the classes within the dependency are used by the project, and all unused classes have been identified and removed. Consequently, there is no class in the API of a specialized dependency that is unused, directly or indirectly, by the project or any other dependency in its dependency tree.}

\revision{Given a project, \deptrim creates a  set of specialized dependencies, such that the dependency tree of the project is composed of dependencies that contain only classes that the project uses.}
Recalling the example in \autoref{fig:jacop-dt-debloated} and \autoref{fig:jacop-dt-specialized}, \jacop uses \np{2} of the \np{2984} classes in \texttt{scala-compiler}. Therefore, \texttt{scala-compiler} could be specialized with respect to \jacop, by removing the \np{2982} unused classes.
\end{definition}

\begin{definition}
\label{def:specialized-dep-tree}
\textbf{Specialized dependency tree: }\normalfont
A specialized dependency tree is a dependency tree where at least one dependency is specialized and the project still correctly builds with that dependency tree. This means that in at least one of the used dependencies, unused classes have been identified and removed. A specialized dependency tree may be one of the following two types: 

\begin{itemize}
    \item \textit{Totally Specialized Tree} (\tst): A dependency tree where all used dependencies are specialized and the project build is successful.
    \item \textit{Partially Specialized Tree} (\pst): A dependency tree with the largest possible number of specialized dependencies, such that the project build is successful.
\end{itemize}
\end{definition}

We discuss our approach for building a project with a \tst or \pst with \deptrim in the following subsections. 
\deptrim identifies unused classes within the non-bloated \compile-scope dependencies of \maven projects, and removes them in order to produce specialized dependencies. 
Using these, \deptrim prepares a specialized dependency tree for the project such that the project still correctly builds. The following subsections explain this technique in detail.   

\subsection{\deptrim}\label{sec:deptrim-challenges}

\begin{figure*}[t]
    \centering
    \noindent\includegraphics[width=1\textwidth]{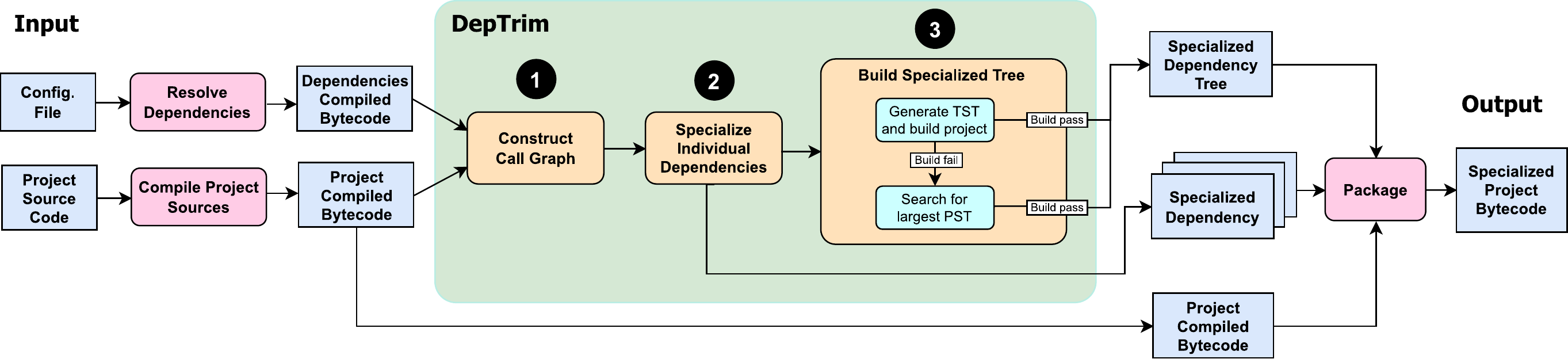} 
    \caption{Overview of the dependency specialization approach implemented in \deptrim. Blue boxes are software artifacts, pink rounded boxes are actions performed by the build engine, and each of the three main phases of \deptrim are indicated within the green rounded box.}
    \label{fig:approach}
    \vspace{-0.25cm}
\end{figure*}

\autoref{fig:approach} illustrates the complete pipeline of the dependency specialization approach implemented in \deptrim.
\deptrim receives as inputs the source code and the \pom file of a Java \maven project. The project must successfully build.
\deptrim outputs three elements: 
\begin{enumerate*} [label=(\roman*)]
\item a specialized version of the \pom file, which removes the bloated dependencies and includes the largest possible number of specialized dependencies that keep the build passing; 
\item the set of specialized dependencies as reduced \jar files;
\item the project compiled from its source can be packaged with the specialized dependencies in order to have a smaller \texttt{jar-with-dependencies} for release and deployment
\end{enumerate*}. 

\deptrim validates that the project still builds correctly with the specialized dependency tree. 
Note that \deptrim only transforms the bytecode of the third-party dependencies, while the original project source and its compiled bytecode remain intact.

As illustrated in \autoref{fig:approach}, the specialization procedure of \deptrim consists of three main phases.
First \ding{202}, \deptrim leverages state-of-the-art Java static bytecode analysis to construct a static call graph of the class members in the third-party dependencies that are reachable from the project binaries.
The completeness of this reachability analysis is critical for the identification of unused third-party classes.
Second \ding{203}, \deptrim transforms the bytecode in the dependencies to remove unused classes.
This task requires integration with the \maven build engine to resolve and deploy the modified dependencies to the local repository.
Finally \ding{204}, \deptrim specializes the dependency tree of the project by modifying its original \pom file.
The modified \pom should preserve the original configurations, except the dependency declarations, which point to the specialized dependencies instead of the original ones.
Moreover, \deptrim must validate that dependency specialization does not break project build. 
We provide more details of these three phases in the following subsections.

\subsubsection{\ding{202} Call Graph Construction}\label{sec:call-graph}

Before it can specialize dependencies, \deptrim determines their API usage, based on static analysis of the project binary. 
To do so, \deptrim constructs a call graph using two inputs: the compiled dependencies as resolved by \maven (Line \autoref{line:algo:dep-resolution} of \autoref{alg:specialization}), and the compiled project sources (Line \autoref{line:algo:compilation}). 
Then, using the bytecode class members of the project as entry points to this graph (Line \autoref{line:algo:call-graph}), \deptrim infers and reports class usage information from the bytecode directly, without loading or initializing classes. The report captures the set of dependencies, classes, and methods that are actually used by the project, \ie, that are reachable via static analysis.
The output of this phase is a data structure that identifies the minimal set of classes in each of the dependencies that are required to build the project.

The collection of accurate and complete call graphs is essential for specialization. 
If a necessary class member is not reachable statically, then \deptrim will consider it as unused and proceed to remove it in a subsequent phase.
To mitigate this limitation, \deptrim relies on state-of-the-art static analysis of Java bytecode to capture invocations between classes, methods, fields, and annotations from the project and its direct and transitive dependencies.
Furthermore, it parses the constant pool of \class files in order to capture dynamic invocations from string literals (\eg, when loading a class using its fully qualified name via reflection).

\subsubsection{\ding{203} Individual Dependency Specialization}

The dependency specialization phase receives the call graph as input to specialize individual dependencies.
During this phase, \deptrim determines which dependencies are bloated (\ie, there is no path from the project bytecode toward any of the class members in the unused dependencies), and removes them from the original \pom (Line \autoref{line:algo:dep-debloating} of \autoref{alg:specialization}).
Next, \deptrim proceeds to remove the unused classes within non-bloated dependencies (Lines \autoref{line:algo:individual-specialization-init} to \autoref{line:algo:individual-specialization-end}). Any dependency \class file that is not present in the call graph is deemed unreachable and removed. 
Note that a Java source file can contain multiple classes, thus resulting in multiple \texttt{class} files after compilation. 
\deptrim considers this case as well by design, as it downloads, unzips, and removes the unused compiled classes directly from the project dependencies at build time (\ie, during the \maven \texttt{package} phase).
Once all the unused \class files in a dependency are removed, \deptrim qualifies the dependency as specialized.
Moreover, to facilitate reuse, \deptrim deploys each specialized dependency in the local \maven repository along with its \pom file and corresponding \texttt{MANIFEST.MF} metadata (Line \autoref{line:algo:specialization-deploy}).

The output of the second phase is a set of specialized \jar files for the dependencies of the project.
These files include all the bytecode and resources that are necessary to be shared and reused by the other packages within the dependency tree. In particular,  \deptrim takes care of keeping the  classes in dependencies that may not be directly instantiated by the project, but are accessible from the used classes in the dependencies, with regard to the project.

\subsubsection{\ding{204} Dependency Tree Specialization}
After specializing each non-bloated dependency, \deptrim produces a specialized version of the project \pom file that removes the bloated dependencies and points to the specialized dependencies instead of their original versions. This results in a \tst or a \pst for the project, as described in Definition~\ref{def:specialized-dep-tree}.

First, \deptrim builds the totally specialized dependency tree (\tst) of the project (Lines \autoref{line:algo:tst-init} to \autoref{line:algo:tst-end} of \autoref{alg:specialization}).
All specialized dependencies replace their original version in the project \pom. Then, in order to validate that the specialization did not remove necessary bytecode, \deptrim builds the project, \ie its sources are compiled and its tests are run. If the build is a \texttt{SUCCESS}, \deptrim returns this \tst.

\begin{algorithm}[t]
\begin{algorithmic}[1]
\caption{Third-party dependency specialization}
\label{alg:specialization}
\small
\renewcommand{\algorithmicrequire}{\textbf{Input:}}
\renewcommand{\algorithmicensure}{\textbf{Output:}}

\REQUIRE $\mathcal{P}_{src}$: Project source code
\REQUIRE  $\mathcal{P}_{obf}$: Project original build file (\pom)
\ENSURE  $\mathcal{P}_\text{\texttt{TST}} \lor \mathcal{P}_\text{\texttt{PST}}$

\textit{\textcolor{olive}{/** Call graph construction **/}}
\STATE $\mathcal{P}_{deps} \gets resolve\_dependencies(\mathcal{P}_{obf})$  \label{line:algo:dep-resolution}
\STATE $\mathcal{P}_{bin} \gets compile(\mathcal{P}_{src})$\label{line:algo:compilation}
\STATE $\mathcal{CG} \gets analyze(\mathcal{P}_{deps}, \mathcal{P}_{bin})$\label{line:algo:call-graph}
\STATE $\mathcal{P}_{dbf} \gets debloat( \mathcal{P}_{obf}, \mathcal{CG})$ \label{line:algo:preparation-end}\label{line:algo:dep-debloating}

\text{\textit{\textcolor{olive}{/** Individual dependency specialization **/}}}
\STATE $\mathcal{P}_{deps\_specialized} \gets \emptyset$\label{line:algo:individual-specialization-init}
\FOR{\textbf{each} $dep \in \mathcal{P}_{dbf}$} \label{line:algo:specialization-iteration-one}
    \STATE $\text{reachable\_classes} \gets analyze(dep, \mathcal{CG})$ \label{line:algo:specialization-reachable}
    \STATE $\text{dep\_specialized }\gets specialize(dep, \text{reachable\_classes})$
    \STATE $\mathcal{P}_{deps\_specialized} \gets \mathcal{P}_{deps\_specialized} \cup \text{dep\_specialized}$
\ENDFOR\label{line:algo:individual-specialization-end}
\STATE
$deploy\_locally(\mathcal{P}_{deps\_specialized})$  \label{line:algo:specialization-deploy} 
\text{\textit{\textcolor{olive}{/** Dependency tree specialization **/}}}
\STATE $\mathcal{P}_\text{\texttt{TST}} \gets \emptyset$ \label{line:algo:tst-init}
\STATE $\mathcal{P}_\text{\texttt{TST}} \gets create\_config\_file(\mathcal{P}_{deps\_specialized})$
\IF{$build(\mathcal{P}_\text{\texttt{TST}}, \mathcal{P}_{bin}) == \text{\texttt{SUCCESS}}$} \label{line:algo:build-tst}
    \RETURN $\mathcal{P}_\text{\texttt{TST}}$ \label{line:algo:tst-end}
\ELSE
    \STATE $\mathcal{P}_\text{\texttt{PST}} \gets \emptyset$\label{line:algo:pst-init} 
    \FOR{\textbf{each} $dep \in \mathcal{P}_{deps\_specialized}$} \label{line:algo:specialization-iteration-two}
    \STATE $\mathcal{P}_{dep} \gets create\_config\_file(dep)$
        \IF{$build(\mathcal{P}_{dep}, \mathcal{P}_{bin}) == \text{\texttt{SUCCESS}}$}
            \STATE $\mathcal{P}_\text{\texttt{PST}} \gets \mathcal{P}_\text{\texttt{PST}} \cup dep$
        \ENDIF
    \ENDFOR
    \RETURN $\mathcal{P}_\text{\texttt{PST}}$ \label{line:algo:specialization-end} \label{line:algo:pst-end}
\ENDIF
\end{algorithmic} 
\end{algorithm}

In cases where the build with the \tst fails, \deptrim proceeds to build the project with one specialized dependency at a time (Lines \autoref{line:algo:pst-init} to \autoref{line:algo:pst-end}).
Thus, rather than attempting to improve the soundness of the static call graph, which is proven to be challenging in Java~\cite{Sui2020}, \deptrim performs an exhaustive search of the dependencies that are candidates for specialization. 
At this step, \deptrim builds as many versions of the dependency tree as there are specialized dependencies, each containing a single specialized dependency.
\deptrim attempts to build the project with each of these single specialized dependency trees.
If the project build is successful, \deptrim marks the dependency as safe to be specialized. 
In case the dependency is not safe to specialize, \deptrim keeps the original dependency entry intact in the specialized \pom file.
Finally, \deptrim constructs a partially specialized dependency tree (\pst) with the union of all the dependencies that are safe to be specialized.
Then, the project is built with this \pst to verify that the build is successful. 
If all build steps pass, \deptrim returns this \pst.

\subsection{Implementation Details}\label{sec:implementation-details}
\deptrim is implemented in Java as a \maven plugin that can be integrated into a project as part of the build pipeline, or be executed directly from the command line.
This design facilitates its integration as part of the projects' CI/CD pipeline, leading to specialized binaries for deployment. 
At its core, \deptrim reuses the state-of-the-art static analysis of \depclean~\cite{ponta2021used}, located in the \texttt{depclean-core} module~\cite{DepcleanCoreModule2023}.
\deptrim adds unique features to this core static Java analyzer by modifying the bytecode within dependencies based on usage information gathered at compilation time, which is different from the complete removal of unused dependencies performed by \depclean.
It uses the ASM Java bytecode analysis library to build a static call graph of \texttt{class} files of the compiled projects and their dependencies. 
The call graph registers usage towards classes, methods, fields, and annotations.
For the deployment of the specialized dependencies, \deptrim relies on the \texttt{deploy-file} goal of the official \texttt{maven-deploy-plugin} from the Apache Software Foundation. 
For dependency analysis and manipulation, \deptrim relies on the \texttt{maven-dependency-plugin}.
\deptrim provides dedicated parameters to target or exclude specific dependencies for specialization, using their identifier and scope.
\deptrim is open-source and reusable from the Maven Central repository. Its source code is publicly available at \href{https://github.com/castor-software/deptrim}{https://github.com/castor-software/deptrim}.

\section{Evaluation}\label{sec:methodology}

Depending on the outcome of specialization, \deptrim potentially removes large portions of the \compile-scope dependencies of the project. 
The output is a specialized distribution that developers should be ready to distribute to users.
\revision{The evaluation described in this section is intended to assess that experience: we run \deptrim on a project, build the project again with specialized dependencies to confirm that its behavior is not negatively impacted (\ie, we use the test suite of the projects as a proxy for checking functional integrity), and evaluate the extent to which our technique is effective.}
Our evaluation is guided by the following research questions:

\newcommand{\RQone}{\revision{What is the impact of removing bloated dependencies on reducing the ratio of third-party code?\xspace}}
\newcommand{\RQtwo}{To what extent can all the used dependencies be specialized and the project built correctly?\xspace}
\newcommand{\RQthree}{How does the number of classes decrease in the dependency tree of the project after specialization?\xspace}
\newcommand{\RQfour}{In what contexts is static dependency specialization  not applicable?\xspace}

\begin{itemize}[leftmargin=26pt]
    \item[\textbf{RQ1}.] \RQone
    \item[\textbf{RQ2}.] \RQtwo
    \item[\textbf{RQ3}.] \RQthree
    \item[\textbf{RQ4}.] \RQfour
\end{itemize}

\subsection{Study Subjects}\label{sec:study-subjects}

\begin{table*}
\centering
\footnotesize
\caption{\revision{Description of the study subjects considered for the evaluation of \deptrim. The table links to the project repository and SHA on GitHub, and lists the number of commits, stars, lines of Java code (LoC), tests, test coverage (\textsc{Cov)}, and the original number of \texttt{compile}-scope dependencies in the project (\#\cd). Also indicated are the number of project classes in each study subject, the total number of classes contributed by {\cd}s, as well as the ratio between them (\ratioo).} 
\textcolor{white}{\texttt{the cake is a lie}}}
\label{tab:descriptive}
\setlength\tabcolsep{3pt}
\begin{tabular}{|l|c|c|r|r|r|r|r|r|r|r|r|} 
\hline
\multirow{2}{*}{\textsc{Project}}                                                                                    & \multirow{2}{*}{\textsc{Module}} & \multirow{2}{*}{\textsc{Commit SHA}}                                                                                                                 & \multicolumn{1}{c|}{\multirow{2}{*}{\textsc{Commits}}} & \multicolumn{1}{c|}{\multirow{2}{*}{\textsc{Stars}}} & \multicolumn{1}{c|}{\multirow{2}{*}{\textsc{LoC}}} & \multicolumn{1}{c|}{\multirow{2}{*}{\textsc{Tests}}} & \multicolumn{1}{c|}{\multirow{2}{*}{\revision{\textsc{Cov. (\%)}}}} & \multicolumn{1}{c|}{\multirow{2}{*}{\textsc{\#CD}}} & \multicolumn{3}{c|}{\textsc{Classes}}                                                                                               \\ 
\cline{10-12}
                                                                                                                                           &                                  &                                                                                                                                                      & \multicolumn{1}{c|}{}                                    & \multicolumn{1}{c|}{}                                  & \multicolumn{1}{c|}{}                                & \multicolumn{1}{c|}{}                                  & \multicolumn{1}{c|}{} & \multicolumn{1}{c|}{}                               & \multicolumn{1}{c|}{\textsc{Project}}  & \multicolumn{1}{c|}{\textsc{CD}}        & \multicolumn{1}{c|}{\ratioo}                \\ 
\hline
\rowcolor[rgb]{0.91,0.918,0.918} 
\href{https://github.com/checkstyle/checkstyle}{\texttt{checkstyle}}                                           & \texttt{--}                      & \texttt{\href{https://github.com/checkstyle/checkstyle/commit/dbeb9024c861ad11b194e40d8c6e08d7e6ec5122}{6ec5122}}        & \np{12066}                   & \np{7455}                  & \np{342795}              & \np{3887}                  & \np{79} & \np{17}                 & \np{863}   & \np{6493}   & \Ratio{863}{6493}      \\
\href{https://github.com/OpenHFT/Chronicle-Map}{\texttt{Chronicle-Map}}                                        & \texttt{--}                      & \texttt{\href{https://github.com/OpenHFT/Chronicle-Map/commit/26e26c132290ad8049c97e0c44eb7f33b63c1c60}{63c1c60}}        & \np{3298}                    & \np{2539}                  & \np{55178}               & \np{1231}                  & \np{49} & \np{35}                 & \np{375}   & \np{7595}   & \Ratio{375}{7595}      \\
\rowcolor[rgb]{0.91,0.918,0.918} 
\href{https://github.com/classgraph/classgraph}{\texttt{classgraph}}                                           & \texttt{--}                      & \texttt{\href{https://github.com/classgraph/classgraph/commit/92b644677964496fb841ba41bed52247f8a24786}{8a24786}}        & \np{5307}                    & \np{2366}                  & \np{32151}               & \np{170}                   & \np{61} & \np{2}                  & \np{261}   & \np{224}    & \Ratio{261}{224}       \\
\href{https://github.com/apache/commons-validator}{\texttt{commons-validator}}                                 & \texttt{--}                      & \texttt{\href{https://github.com/apache/commons-validator/commit/f9bb21748a9f9c50fbc31862de25ed49433ecc88}{33ecc88}}     & \np{1742}                    & \np{155}                   & \np{16781}               & \np{576}                   & \np{85} & \np{4}                  & \np{64}    & \np{780}    & \Ratio{64}{780}        \\
\rowcolor[rgb]{0.91,0.918,0.918} 
\href{https://github.com/stanfordnlp/CoreNLP}{\texttt{CoreNLP}}                                                & \texttt{--}                      & \texttt{\href{https://github.com/checkstyle/checkstyle/commit/f7782ff5f235584b0fc559f266961b5ab013556a}{013556a}}        & \np{17012}                   & \np{8802}                  & \np{605561}              & \np{1374}                  & \np{18} & \np{32}                 & \np{3932}  & \np{9121}   & \Ratio{3932}{9121}     \\
\href{https://github.com/apache/flink}{\texttt{flink}}                                                         & \texttt{flink-java}              & \texttt{\href{https://github.com/apache/flink/commit/c41c8e5cfab683da8135d6c822693ef851d6e2b7}{1d6e2b7}}                 & \np{32667}                   & \np{20488}                 & \np{36455}               & \np{836}                   & \np{56} & \np{16}                 & \np{277}   & \np{6175}   & \Ratio{277}{6175}      \\
\rowcolor[rgb]{0.91,0.918,0.918} 
\href{https://github.com/graphhopper/graphhopper}{\texttt{graphhopper}}                                        & \texttt{core}                    & \texttt{\href{https://github.com/graphhopper/graphhopper/commit/6d3da37960f56aa6b9c4b1ffd77f70ebebff8747}{bff8747}}      & \np{6211}                    & \np{3978}                  & \np{66119}               & \np{2460}                  & \np{81} & \np{18}                 & \np{631}   & \np{5474}   & \Ratio{631}{5474}      \\
\href{https://github.com/google/guice}{\texttt{guice}}                                                         & \texttt{core}                    & \texttt{\href{https://github.com/google/guice/commit/b0ff10c8ec8911137451623a333d6daa65f73d8a}{5f73d8a}}                 & \np{2026}                    & \np{11730}                 & \np{49697}               & \np{979}                   & \np{90} & \np{10}                 & \np{460}   & \np{2474}   & \Ratio{460}{2474}      \\
\rowcolor[rgb]{0.91,0.918,0.918} 
\href{https://github.com/helidon-io/helidon}{\texttt{helidon-io}}                                              & \texttt{openapi}                 & \texttt{\href{https://github.com/helidon-io/helidon/commit/99cf5add9b4049581f08aae9eddaf0280070f2bb}{070f2bb}}           & \np{2707}                    & \np{2929}                  & \np{7729}                & \np{30}                    & \np{76} & \np{36}                 & \np{32}    & \np{4002}   & \Ratio{32}{4002}       \\
\href{https://github.com/apache/httpcomponents-client/tree/master/httpclient5}{\texttt{httpcomponents}} & \texttt{httpclient5}             & \texttt{\href{https://github.com/apache/httpcomponents-client/commit/d8f702fb4d44c746bb0edf00643aa7139cb8bdf7}{cb8bdf7}} & \np{3424}                    & \np{1269}                  & \np{42920}               & \np{669}                   & \np{50} & \np{5}                  & \np{493}   & \np{1153}   & \Ratio{493}{1153}      \\
\rowcolor[rgb]{0.91,0.918,0.918} 
\href{https://github.com/immutables/immutables}{\texttt{immutables}}                                           & \texttt{gson}                    & \texttt{\href{https://github.com/immutables/immutables/commit/6e192030320eaf7a8b5f146c39ae5a17b413aa37}{413aa37}}        & \np{2588}                    & \np{3218}                  & \np{16448}               & \np{37}                    & \np{60} & \np{2}                  & \np{31}    & \np{307}    & \Ratio{31}{307}        \\
\href{https://github.com/radsz/jacop}{\texttt{jacop}}                                                          & \texttt{--}                      & \texttt{\href{https://github.com/radsz/jacop/commit/1a395e6add22caf79590fe9d1b2223bfb6ed0cd0}{6ed0cd0}}                  & \np{1302}                    & \np{202}                   & \np{93170}               & \np{210}                   & \np{34} & \np{9}                  & \np{833}   & \np{8487}   & \Ratio{833}{8487}      \\
\rowcolor[rgb]{0.91,0.918,0.918} 
\href{https://github.com/DiUS/java-faker}{\texttt{java-faker}}                                                 & \texttt{--}                      & \texttt{\href{https://github.com/DiUS/java-faker/commit/e23d6067c8f83b335a037d24e6107a37eb0b9e6e}{b0b9e6e}}              & \np{834}                     & \np{3914}                  & \np{8429}                & \np{579}                   & \np{98} & \np{4}                  & \np{107}   & \np{503}    & \Ratio{107}{503}       \\
\href{https://github.com/jcabi/jcabi-github}{\texttt{jcabi-github}}                                            & \texttt{--}                      & \texttt{\href{https://github.com/jcabi/jcabi-github/commit/02f3ab93156349c2f66989ac675bd6292462d724}{462d724}}           & \np{2764}                    & \np{276}                   & \np{33542}               & \np{684}                   & \np{44} & \np{20}                 & \np{312}   & \np{3921}   & \Ratio{312}{3921}      \\
\rowcolor[rgb]{0.91,0.918,0.918} 
\href{https://github.com/google/jimfs}{\texttt{jimfs}}                                                         & \texttt{jimfs}                   & \texttt{\href{https://github.com/google/jimfs/commit/3bc54fae2feb218dcf5427d2626720fc09ef38d1}{9ef38d1}}                 & \np{508}                     & \np{2234}                  & \np{15558}               & \np{5834}                  & \np{91} & \np{9}                  & \np{124}   & \np{3560}   & \Ratio{124}{3560}      \\
\href{https://github.com/jooby-project/jooby/tree/3.x/jooby}{\texttt{jooby}}                                   & \texttt{jooby}                   & \texttt{\href{https://github.com/jooby-project/jooby/commit/4d7be54dad429b5aeb5266387df14b0781c78357}{1c78357}}          & \np{4702}                    & \np{1523}                  & \np{20154}               & \np{122}                   & \np{31} & \np{22}                 & \np{320}   & \np{6945}   & \Ratio{320}{6945}      \\
\rowcolor[rgb]{0.91,0.918,0.918} 
\href{https://github.com/lettuce-io/lettuce-core}{\texttt{lettuce}}                                            & \texttt{core}                    & \texttt{\href{https://github.com/lettuce-io/lettuce-core/commit/77c4ea587fdba73f688c95a481d2743b6fc94fcb}{fc94fcb}}      & \np{2280}                    & \np{4861}                  & \np{89468}               & \np{2600}                  & \np{42} & \np{44}                 & \np{1302}  & \np{10364}  & \Ratio{1302}{10364}    \\
\href{https://github.com/modelmapper/modelmapper}{\texttt{modelmapper}}                                        & \texttt{core}                    & \texttt{\href{https://github.com/modelmapper/modelmapper/commit/894c658609f47f817ce20e41f6e87e1f803663ee}{03663ee}}      & \np{721}                     & \np{2090}                  & \np{21769}               & \np{618}                   & \np{84} & \np{6}                  & \np{210}   & \np{2700}   & \Ratio{210}{2700}      \\
\rowcolor[rgb]{0.91,0.918,0.918} 
\href{https://github.com/mybatis/mybatis-3}{\texttt{mybatis-3}}                                                & \texttt{--}                      & \texttt{\href{https://github.com/mybatis/mybatis-3/commit/c195f12808a88a1ee245dc86d9c1621042655970}{2655970}}            & \np{4436}                    & \np{18065}                 & \np{61849}               & \np{1699}                  & \np{86} & \np{8}                  & \np{480}   & \np{1345}   & \Ratio{480}{1345}      \\
\href{https://github.com/LibrePDF/OpenPDF}{\texttt{OpenPDF}}                                                   & \texttt{--}                      & \texttt{\href{https://github.com/LibrePDF/OpenPDF/commit/0c9c4ca393b01444b7cb13bb1d12da202bd0d458}{bd0d458}}             & \np{1296}                    & \np{2573}                  & \np{76397}               & \np{35}                    & \np{29} & \np{35}                 & \np{484}   & \np{17512}  & \Ratio{484}{17512}     \\
\rowcolor[rgb]{0.91,0.918,0.918} 
\href{https://github.com/apache/pdfbox/tree/trunk/pdfbox}{\texttt{pdfbox}}                                     & \texttt{pdfbox}                  & \texttt{\href{https://github.com/apache/pdfbox/commit/e72963ca5b283a87828ee731cd85c0b6baf1ff57}{af1ff57}}                & \np{11147}                   & \np{1852}                  & \np{97175}               & \np{654}                   & \np{79} & \np{7}                  & \np{754}   & \np{6836}   & \Ratio{754}{6836}      \\
\href{https://github.com/pf4j/pf4j}{\texttt{pf4j}}                                                             & \texttt{--}                      & \texttt{\href{https://github.com/pf4j/pf4j/commit/efaed93c10dd9d114335e2a344e8bca04fd00c63}{fd00c63}}                    & \np{692}                     & \np{1901}                  & \np{7199}                & \np{151}                   & \np{73} & \np{3}                  & \np{93}    & \np{115}    & \Ratio{93}{115}        \\
\rowcolor[rgb]{0.91,0.918,0.918} 
\href{https://github.com/Sayi/poi-tl}{\texttt{poi-tl}}                                                         & \texttt{--}                      & \texttt{\href{https://github.com/Sayi/poi-tl/commit/5d5004311a406b7d5843be76322bf208071b5969}{71b5969}}                  & \np{732}                     & \np{3063}                  & \np{20882}               & \np{125}                   & \np{79} & \np{36}                 & \np{255}   & \np{12143}  & \Ratio{255}{12143}     \\
\href{https://github.com/Col-E/Recaf}{\texttt{Recaf}}                                                          & \texttt{--}                      & \texttt{\href{https://github.com/Col-E/Recaf/commit/c66f23801493bd866db757b0594c1fceaa30dce0}{a30dce0}}                  & \np{2275}                    & \np{4530}                  & \np{31277}               & \np{274}                   & \np{35} & \np{56}                 & \np{538}   & \np{10769}  & \Ratio{538}{10769}     \\
\rowcolor[rgb]{0.91,0.918,0.918} 
\href{https://github.com/JakeWharton/RxRelay}{\texttt{RxRelay}}                                                & \texttt{--}                      & \texttt{\href{https://github.com/JakeWharton/RxRelay/commit/e9fc1586192ca1ecdbc41ae39036cbf0d09428b5}{09428b5}}          & \np{81}                      & \np{2473}                  & \np{2405}                & \np{64}                    & \np{93} & \np{2}                  & \np{16}    & \np{1758}   & \Ratio{16}{1758}       \\
\href{https://github.com/scribejava/scribejava}{\texttt{scribejava}}                                           & \texttt{--}                      & \texttt{\href{https://github.com/scribejava/commit/466f37c6faf5a9a2de9c87ae1bce71b617a6185b}{7a6185b}}                   & \np{1259}                    & \np{5317}                  & \np{5769}                & \np{82}                    & \np{45} & \np{8}                  & \np{116}   & \np{1278}   & \Ratio{116}{1278}      \\
\rowcolor[rgb]{0.91,0.918,0.918} 
\href{https://github.com/jtablesaw/tablesaw/tree/master/json}{\texttt{tablesaw}}                               & \texttt{json}                    & \texttt{\href{https://github.com/jtablesaw/tablesaw/commit/05823f66246ea191e62ad0658d2fed0b080d5334}{80d5334}}           & \np{2501}                    & \np{3101}                  & \np{508}                 & \np{9}                     & \np{75} & \np{9}                  & \np{7}     & \np{1447}   & \Ratio{7}{1447}        \\
\href{https://github.com/apache/tika/tree/main/tika-core}{\texttt{tika}}                                       & \texttt{tika-core}               & \texttt{\href{https://github.com/apache/tika/commit/41319f3c294b13de5342a80570b4540f7dd04a3e}{dd04a3e}}                  & \np{6823}                    & \np{1584}                  & \np{32388}               & \np{305}                   & \np{50} & \np{2}                  & \np{435}   & \np{253}    & \Ratio{435}{253}       \\
\rowcolor[rgb]{0.91,0.918,0.918} 
\href{https://github.com/undertow-io/undertow/tree/master/core}{\texttt{undertow}}                             & \texttt{core}                    & \texttt{\href{https://github.com/undertow-io/undertow/commit/56c91f129b1c2a55cf3287836cc68c80acce54c6}{cce54c6}}         & \np{5517}                    & \np{3284}                  & \np{106711}              & \np{682}                   & \np{59} & \np{5}                  & \np{1581}  & \np{742}    & \Ratio{1581}{742}      \\
\href{https://github.com/FasterXML/woodstox}{\texttt{woodstox}}                                                & \texttt{--}                      & \texttt{\href{https://github.com/FasterXML/woodstox/commit/e8f00401bebd103f62d51383ef53da2cd58bd89e}{58bd89e}}           & \np{325}                     & \np{180}                   & \np{60476}               & \np{868}                   & \np{67} & \np{5}                  & \np{208}   & \np{864}    & \Ratio{208}{864}       \\
\hline
\textsc{Total}                                                                                                 & \np{14}                          & \np{30}                                                                                                                  & \np{139243}                  & \np{127952}                & \np{2056960}             & \np{27844}                 & \textsc{(Med.)} \np{64} & \np{467}                & \np{15594} & \np{135343} & \Ratio{15594}{135343}  \\
\hline
\end{tabular}
\end{table*}


We evaluate \deptrim with \np{30} open-source projects collected from two data sources. 
The first source is the dataset of single-module Java projects made available by Durieux~\etal~\cite{Durieux2021}. 
This dataset contains \np{395} popular projects that build successfully with \maven, \ie all their tests pass, and a compiled artifact is produced as a result of the build.
We analyze the dependency tree of the projects in this dataset and select those that have at least one \compile-scope dependency.
This results in \np{13} projects.
Additionally, we derive a second set of projects through the advanced search feature of GitHub.
We filter repositories with a \pom file and rank the resulting Java \maven projects in descending order according to the number of stars.
We rely on the number of stars as a proxy for popularity~\cite{borges2018s}.
\revision{Then, we curate these projects to have 17 projects that meet the following criteria: (i) build successfully with \maven, \ie compile and all their tests pass,  (ii) declare at least one \compile-scope dependency, and (iii) have at least one test executed by the \texttt{maven-surefire-plugin}.
We build the projects at least two times to avoid including projects with flaky tests. 
At the end of this curation process, we have a set of \np{30} study subjects with at least one \compile-scope dependency, and an executable test suite with tests that pass.}


\autoref{tab:descriptive} presents descriptive statistics for the \np{30} study subjects.
For multi-module projects, we specify the \textsc{Module} we use for our experiments. 
Furthermore, we link to the \textsc{Commit SHA} of the  version that we consider for the evaluation.
The explicit documentation of \textsc{Project}, \textsc{Module}, and \textsc{Commit SHA} ensure the reproducibility of our evaluation.
The projects are well-known in the Java community and have between \np{155} and \np{20488} \textsc{Stars}, for \texttt{commons-validator} and  \texttt{flink} respectively. The median number of stars is \np{2751}. 
\texttt{flink} also has the maximum number of \textsc{Commits}, at \np{32667}, while the median number of commits across the study subjects is \np{2544}.
Next, we report the number of lines of Java code (\textsc{LoC}) in each project, computed with the Unix command \texttt{cloc}.
In total, the projects have more than \np[M]{2} \textsc{LoC}.
The two projects with the largest number of lines of code are \texttt{CoreNLP} (\np{605561}) and \texttt{checkstyle} (\np{342795}), while the median \textsc{LoC} across the projects is \np{32965}.
In the \textsc{Tests}  column we give the number of tests executed by the official \texttt{maven-surefire-plugin} in the projects. The median number of tests is \np{599}. The two projects with the most tests are \texttt{jimfs} (\np{5834}) and \texttt{checkstyle} (\np{3887}).
\revision{In the \textsc{Cov.}  column we provide the code coverage of the test suite of the project, as measured with \texttt{JaCoCo}\footnote{\url{https://www.eclemma.org/jacoco/}}.
The median coverage for the $30$ study subjects is $64$\%.
When we study one specific module of a multi-module project, the \textsc{LoC}, \textsc{Tests}, and \textsc{Cov.} numbers are for the specific module under study.}

The last 4 columns of \autoref{tab:descriptive} provide dependency-specific information. First, the number of \compile-scope dependencies as resolved by \maven (\textsc{\#CD}). There are \np{467} \compile-scope dependencies across the \np{30} projects, with a  median number of  \np{9} CDs and at least \np{2} CDs in each project. The maximum number of \compile-scope dependencies  is \np{56}, in \texttt{Recaf}.
The following columns present the number of \textsc{classes} that are written by the developers of the \textsc{project}, and the number of third-party classes that come from its \compile-scope dependencies (\textsc{CD}). 
The bytecode of each of these classes is analyzed by \deptrim in order to construct a static call graph of APIs usages between the projects and dependencies, as described in \autoref{sec:call-graph}.
In total, \deptrim analyzes the bytecode of \np{15594} project classes, and \np{135343} classes from third-party dependencies.
\texttt{CoreNLP} has \np{3932} project classes, the maximum in the dataset. The largest number of third-party classes is \np{17512}, in \texttt{OpenPDF}.
In the last column of the table, we present the dependency classes to project classes ratio  (\ratioo in \autoref{eq:ratio-original}). 

\begin{equation} \label{eq:ratio-original}
\scriptsize
\textsc{Ratio}_\mathcal{O} = \frac{\textsc{\#CD Classes}}{\textsc{\#Project Classes}} 
\end{equation}

We find that, for \np{27} of these \np{30} notable projects, most of the code actually belongs to third-party dependencies. 
In fact, this ratio is as high as \np{206.7}$\times$ for the project \texttt{tablesaw}. 
Across our dataset, the ratio of the project classes to the dependency classes is \np{8.7}$\times$.

Recalling the example of \jacop introduced in \autoref{fig:jacop-dt}, the corresponding row in  \autoref{tab:descriptive} reads as follows: we select its latest release for our evaluation (SHA \texttt{\href{https://github.com/radsz/jacop/commit/1a395e6add22caf79590fe9d1b2223bfb6ed0cd0}{6ed0cd0}}), which has \np{1302} commits, \np{93170} lines of Java code, \np{210} tests, and has been starred by \np{202} users on GitHub. When \jacop is compiled, the number of classes from \jacop is \np{833}. On the other hand, its \np{9} \compile-scope dependencies contribute \np{10.2}$\times$ more classes (\ie, \np{8487})  compared to the classes in the project (\ie, \nbClassesInJacopProject).

\subsection{Protocol for RQ1}\label{sec:protocol-rq1}

With this research question, we quantify the potential for dependency specialization in the \np{30} projects described in \autoref{tab:descriptive}.
In order to do so, we use \depclean to identify and remove bloated dependencies from each project, ensuring that the project still builds.
We report the number of \compile-scope dependencies that are non-bloated, denoted as \nbcd.
Next, we present the total number of classes removed through dependency debloating (\textsc{Classes Removed}), and compute the ratio (\ratiod) between the remaining dependency classes and the project classes (per \autoref{eq:ratio-debloated}).
This data provides quantitative insights regarding the impact of dependency debloating to reduce the share of third-party code, and on the opportunity to reduce this share further via dependency specialization.

\begin{equation} \label{eq:ratio-debloated}
 \scriptsize
\textsc{Ratio$_\mathcal{D}$} = \frac{\textsc{\#CD Classes} - \textsc{\#Classes Removed by DepClean}}{\textsc{\#Project Classes}} 
\end{equation}

\subsection{Protocol for RQ2}\label{sec:protocol-rq2}

In order to answer RQ2, we attach \deptrim to the \maven build lifecycle of each of our study subjects.
\deptrim is implemented as a \maven plugin, which facilitates this integration, as described in \autoref{sec:implementation-details}.
This means that the non-bloated \compile-scope dependencies in the dependency tree of each project are resolved, specialized, and deployed to the local \maven repository. 
\deptrim then attempts to build the project, \ie, compile it and run its tests, with the goal of preparing a specialized dependency tree with the maximum number of specialized dependencies.
Per \autoref{alg:specialization}, \deptrim constructs either a totally specialized tree (\tst), or a partially specialized tree (\pst) that includes the largest number of specialized dependencies that preserve the build correctness. 
For each project, we report whether it builds with a \tst. If it does not, we report the number of dependencies that \deptrim successfully specializes to prepare a \pst (through the metric \textsc{NBCD Specialized}). 
The findings from this research question highlight the applicability of \deptrim on real-world \maven projects, and its ability to prepare minimal versions of these projects, by removing unused classes within non-bloated dependencies while passing the build.

\subsection{Protocol for RQ3}\label{sec:protocol-rq3}

After building each project successfully with a \tst or a \pst in RQ2, we report the total number of classes that are removed by \deptrim through the specialization of its non-bloated \compile-scope dependencies (as \textsc{Classes Removed}). 
We also report the ratio of the remaining number of specialized dependency classes to the number of project classes (\ratios in \autoref{eq:ratio-specialized}).
We compare \ratios with \ratioo, \ie, we evaluate the reduction in the original ratio after specialization. 
This research question demonstrates the practical advantages of dependency specialization with \deptrim, specifically the reduction in the original proportion of third-party classes within the compiled project binary.

\begin{equation} \label{eq:ratio-specialized}
\scriptsize
\textsc{Ratio$_\mathcal{S}$} = \frac{\textsc{\#NBCD Classes} - \textsc{\#Classes Removed by DepTrim}}{\textsc{\#Project Classes}} 
\end{equation}

\subsection{Protocol for RQ4}\label{sec:protocol-rq4}

While processing each project, \deptrim records the project build logs after dependency specialization (RQ2), as well as the number of classes removed from each non-bloated dependency (RQ3). We derive the answer for RQ4 by analyzing these logs.
In some cases, all the classes in a non-bloated dependency are used by the project, leaving no room for specialization. We refer to such a dependency as a totally used dependency (\textsc{TUD}). We report the number of TUDs for each project, where \deptrim is not applicable by design.
Another situation where \deptrim is not applicable is when a project uses dynamic features to access dependency classes (\autoref{sec:call-graph}). While computing the \pst for RQ2, \deptrim builds the project multiple times, each time with a single specialized dependency. In case of a failure when building the project  with a specialized dependency, we report a compilation error or a test failure.
For the assessment of the compilation results, we rely on the official \texttt{maven-compiler-plugin}.
We consider the execution of the test suite to fail if there is at least one test reported within the sets of \texttt{Failures} or \texttt{Errors}, as reported by the official \texttt{maven-surefire-plugin}.
With this research question, we gain insights regarding the existing challenges of dependency specialization with \deptrim.
More generally, it contributes to the understanding of the limitations of static analysis with respect to specializing dependencies, in view of the dynamic features of Java.

\subsection{Evaluation Framework}\label{sec:evaluation-framework}

In order to run our experiments, we have designed a fully automated framework that orchestrates the execution of \deptrim, the creation of specialized dependency trees, the building of the projects with the specialized dependency trees, as well as the collection and processing of data to answer our research questions.
Since \deptrim is implemented as a \maven plugin, it integrates within the \maven build lifecycle and executes during the \texttt{package} phase.
The execution was performed on a virtual machine running Ubuntu Server with \np[cores]{16} of CPU and \np[GB]{32} of RAM.
It took one week to execute the complete experiment with the \np{30} study subjects. 
This execution  time is essentially due to multiple executions of the large test suites of  our  subjects: once with the original project; once after debloating dependencies with \depclean; once with the \tst generated by \deptrim, and if we generate a \pst for a project, we run  the test suite once with each individually specialized tree and once with the final \pst. 
The execution framework is publicly available on GitHub at \href{https://github.com/castor-software/deptrim-experiments}{castor-software/deptrim-experiments}, and the raw data obtained from the complete execution is available on Zenodo at \href{https://doi.org/10.5281/zenodo.7613554}{10.5281/zenodo.7613554}.

\section{Experimental Results}\label{sec:results}

This section presents the results from our evaluation of \deptrim with the \np{30} Java projects described in \autoref{sec:study-subjects}. We evaluate the effectiveness of \deptrim in automatically specializing the dependency tree of these projects. The answers to the four RQs are summarized in \autoref{tab:deptrim}.

\subsection{RQ1: \RQone}\label{sec:results-rq1}

\begin{table*}
\footnotesize
\centering
\caption{Results from the evaluation of \deptrim with the case studies described in \autoref{tab:descriptive}. For RQ1, the table indicates the number of non-bloated \compile-scope dependencies (\nbcd). These are the dependencies that are identified as used in the project by \depclean. The number of classes removed via debloating is listed in the column \textsc{Classes Removed}.
\ratiod represents the number of remaining third-party classes after debloating over the number of classes in the project.
For RQ2, we highlight whether \deptrim builds a project with a \tst or a \pst, as well as the number of specialized NBCDs contained in the built project (\textsc{NBCD Specialized}).
RQ3 presents the reduction in the number of classes in the NBCDs as a result of specialization with \deptrim, and correspondingly, in the \ratios of the third-party classes to project classes.
For RQ4, we report the three cases where specialization with \deptrim is not applicable: (i) if all the classes in a non-bloated dependency are totally used by the project (\tud); (ii) if specializing a dependency causes a compilation error (\textsc{Comp. Error}) during project build; and (iii) if a test fails when building a project with a specialized dependency (\textsc{Test Fail.}). \textcolor{white}{\href{https://github.com/DiUS/java-faker/blob/master/src/main/resources/en/hitchhikers\_guide\_to\_the\_galaxy.yml\#L41}{DON'T PANIC}}}
\label{tab:deptrim}
\setlength\tabcolsep{2pt}
\begin{tabular}{|l||r|r|r||c|c|r||r|r||r|r|r|}
\hline
\multirow{5}{*}{\textsc{Project}} &
\multicolumn{3}{c||}{{\cellcolor[rgb]{0.961,0.957,0.996}}RQ1} &
\multicolumn{3}{c||}{{\cellcolor[rgb]{0.988,0.996,0.922}}RQ2} &
\multicolumn{2}{c||}{{\cellcolor[rgb]{0.886,0.992,0.988}}RQ3} &
\multicolumn{3}{c|}{{\cellcolor[rgb]{0.882,1,0.882}}RQ4} \\
\hhline{|~||---||---||--||---|} &
{\cellcolor[rgb]{0.961,0.957,0.996}} \textsc{NBCD} &
{\cellcolor[rgb]{0.961,0.957,0.996}} {\begin{tabular}[c]{@{}c@{}}\textsc{Classes} \\ \textsc{Removed} \\ \textsc{(Debloating)} \end{tabular}} &
{\cellcolor[rgb]{0.961,0.957,0.996}} \textsc{\ratiod} &
{\cellcolor[rgb]{0.988,0.996,0.922}} \textsc{\tst} &
{\cellcolor[rgb]{0.988,0.996,0.922}} \textsc{\pst} &
{\cellcolor[rgb]{0.988,0.996,0.922}} {\begin{tabular}[c]{@{}c@{}}\textsc{NBCD} \\ \textsc{Specialized} \end{tabular}} &
{\cellcolor[rgb]{0.886,0.992,0.988}} {\begin{tabular}[c]{@{}c@{}}\textsc{Classes} \\ \textsc{Removed} \\ \textsc{(Specialization)} \end{tabular}} &
{\cellcolor[rgb]{0.886,0.992,0.988}} \textsc{\ratios} &
{\cellcolor[rgb]{0.882,1,0.882}} \textsc{TUD} &
{\cellcolor[rgb]{0.882,1,0.882}} {\begin{tabular}[c]{@{}c@{}}\textsc{Comp.} \\ \textsc{Error} \end{tabular}} &
{\cellcolor[rgb]{0.882,1,0.882}} {\begin{tabular}[c]{@{}c@{}}\textsc{Test} \\ \textsc{Fail.} \end{tabular}} \\
\hline
\rowcolor[rgb]{0.91,0.918,0.918} \texttt{checkstyle} &
\np{15}/\np{17} &
\ChartSmall[p]{2}{6493} &
\Ratio{863}{6491} &
\textcolor{red}{\ding{55}} &
\textcolor{teal}{\ding{51}} &
\np{12}/\np{15} &
\ChartSmall[p]{1015}{6491} &
\Ratio{863}{5476} &
\np{0}/\np{15} &
\np{2}/\np{3} &
\np{1}/\np{3} \\
\texttt{Chronicle-Map} &
{\cellcolor[rgb]{0.996,1,1}}\np{28}/\np{35} &
{\cellcolor[rgb]{0.996,1,1}}\ChartSmall[p]{373}{7595} &
{\cellcolor[rgb]{0.996,1,1}}\Ratio{375}{7222} &
{\cellcolor[rgb]{0.996,1,1}}\textcolor{red}{\ding{55}} &
{\cellcolor[rgb]{0.996,1,1}}\textcolor{teal}{\ding{51}} &
{\cellcolor[rgb]{0.996,1,1}}\np{22}/\np{28} &
{\cellcolor[rgb]{0.996,1,1}}\ChartSmall[p]{2573}{7317} &
{\cellcolor[rgb]{0.996,1,1}}\Ratio{375}{4649} &
{\cellcolor[rgb]{0.996,1,1}}\np{4}/\np{28} &
{\cellcolor[rgb]{0.996,1,1}}\np{1}/\np{2} &
{\cellcolor[rgb]{0.996,1,1}}\np{1}/\np{2} \\
\rowcolor[rgb]{0.91,0.918,0.918} \texttt{classgraph} &
\np{2}/\np{2} &
\ChartSmall[p]{0}{224} &
\Ratio{261}{224} &
\textcolor{teal}{\ding{51}} &
&
\np{2}/\np{2} &
\ChartSmall[p]{10}{224} &
\Ratio{261}{214} &
\np{0}/\np{2} &
-- &
-- \\
\texttt{commons-validator} &
{\cellcolor[rgb]{0.996,1,1}}\np{4}/\np{4} &
{\cellcolor[rgb]{0.996,1,1}}\ChartSmall[p]{0}{780} &
{\cellcolor[rgb]{0.996,1,1}}\Ratio{64}{780} &
{\cellcolor[rgb]{0.996,1,1}}\textcolor{teal}{\ding{51}} &
{\cellcolor[rgb]{0.996,1,1}} &
{\cellcolor[rgb]{0.996,1,1}}\np{4}/\np{4} &
{\cellcolor[rgb]{0.996,1,1}}\ChartSmall[p]{625}{780} &
{\cellcolor[rgb]{0.996,1,1}}\Ratio{64}{155} &
{\cellcolor[rgb]{0.996,1,1}}\np{0}/\np{4} &
{\cellcolor[rgb]{0.996,1,1}}-- &
{\cellcolor[rgb]{0.996,1,1}}-- \\
\rowcolor[rgb]{0.91,0.918,0.918} \texttt{CoreNLP} &
\np{30}/\np{32} &
\ChartSmall[p]{364}{9121} &
\Ratio{3932}{8757} &
\textcolor{teal}{\ding{51}} &
&
\np{29}/\np{30} &
\ChartSmall[p]{3648}{8757} &
\Ratio{3932}{5109} &
\np{1}/\np{30} &
-- &
-- \\
\texttt{flink} &
{\cellcolor[rgb]{0.996,1,1}}\np{15}/\np{16} &
{\cellcolor[rgb]{0.996,1,1}}\ChartSmall[p]{0}{6175} &
{\cellcolor[rgb]{0.996,1,1}}\Ratio{277}{6175} &
{\cellcolor[rgb]{0.996,1,1}}\textcolor{red}{\ding{55}} &
{\cellcolor[rgb]{0.996,1,1}}\textcolor{teal}{\ding{51}} &
{\cellcolor[rgb]{0.996,1,1}}\np{12}/\np{15} &
{\cellcolor[rgb]{0.996,1,1}}\ChartSmall[p]{2594}{6175} &
{\cellcolor[rgb]{0.996,1,1}}\Ratio{277}{3581} &
{\cellcolor[rgb]{0.996,1,1}}\np{1}/\np{15} &
{\cellcolor[rgb]{0.996,1,1}}\np{1}/\np{2} &
{\cellcolor[rgb]{0.996,1,1}}\np{1}/\np{2} \\
\rowcolor[rgb]{0.91,0.918,0.918} \texttt{graphhopper} &
\np{13}/\np{18} &
\ChartSmall[p]{1309}{5474} &
\Ratio{631}{4165} &
\textcolor{red}{\ding{55}} &
\textcolor{teal}{\ding{51}} &
\np{12}/\np{13} &
\ChartSmall[p]{1661}{4165} &
\Ratio{631}{2504} &
\np{0}/\np{13} &
\np{1}/\np{1} &
\np{0}/\np{1} \\
\texttt{guice} &
{\cellcolor[rgb]{0.996,1,1}}\np{9}/\np{10} &
{\cellcolor[rgb]{0.996,1,1}}\ChartSmall[p]{0}{2474} &
{\cellcolor[rgb]{0.996,1,1}}\Ratio{460}{2474} &
{\cellcolor[rgb]{0.996,1,1}}\textcolor{teal}{\ding{51}} &
{\cellcolor[rgb]{0.996,1,1}} &
{\cellcolor[rgb]{0.996,1,1}}\np{7}/\np{9} &
{\cellcolor[rgb]{0.996,1,1}}\ChartSmall[p]{1327}{2474} &
{\cellcolor[rgb]{0.996,1,1}}\Ratio{460}{1147} &
{\cellcolor[rgb]{0.996,1,1}}\np{2}/\np{9} &
{\cellcolor[rgb]{0.996,1,1}}-- &
{\cellcolor[rgb]{0.996,1,1}}-- \\
\rowcolor[rgb]{0.91,0.918,0.918} \texttt{helidon-io} &
\np{34}/\np{36} &
\ChartSmall[p]{38}{4002} &
\Ratio{32}{3964} &
\textcolor{red}{\ding{55}} &
\textcolor{teal}{\ding{51}} &
\np{32}/\np{34} &
\ChartSmall[p]{987}{3964} &
\Ratio{32}{2977} &
\np{1}/\np{34} &
\np{1}/\np{1} &
\np{0}/\np{1} \\
\texttt{httpcomponents} &
{\cellcolor[rgb]{0.996,1,1}}\np{5}/\np{5} &
{\cellcolor[rgb]{0.996,1,1}}\ChartSmall[p]{0}{1153} &
{\cellcolor[rgb]{0.996,1,1}}\Ratio{493}{1153} &
{\cellcolor[rgb]{0.996,1,1}}\textcolor{teal}{\ding{51}} &
{\cellcolor[rgb]{0.996,1,1}} &
{\cellcolor[rgb]{0.996,1,1}}\np{4}/\np{5} &
{\cellcolor[rgb]{0.996,1,1}}\ChartSmall[p]{432}{1153} &
{\cellcolor[rgb]{0.996,1,1}}\Ratio{493}{721} &
{\cellcolor[rgb]{0.996,1,1}}\np{1}/\np{5} &
{\cellcolor[rgb]{0.996,1,1}}-- &
{\cellcolor[rgb]{0.996,1,1}}-- \\
\rowcolor[rgb]{0.91,0.918,0.918} \texttt{immutables} &
\np{2}/\np{2} &
\ChartSmall[p]{0}{307} &
\Ratio{31}{307} &
\textcolor{teal}{\ding{51}} &
&
\np{2}/\np{2} &
\ChartSmall[p]{48}{307} &
\Ratio{31}{259} &
\np{0}/\np{2} &
-- &
-- \\
\texttt{jacop} &
{\cellcolor[rgb]{0.996,1,1}}\np{4}/\np{9} &
{\cellcolor[rgb]{0.996,1,1}}\ChartSmall[p]{504}{8487} &
{\cellcolor[rgb]{0.996,1,1}}\Ratio{833}{7983} &
{\cellcolor[rgb]{0.996,1,1}}\textcolor{red}{\ding{55}} &
{\cellcolor[rgb]{0.996,1,1}}\textcolor{teal}{\ding{51}} &
{\cellcolor[rgb]{0.996,1,1}}\np{3}/\np{4} &
{\cellcolor[rgb]{0.996,1,1}}\ChartSmall[p]{5704}{7983} &
{\cellcolor[rgb]{0.996,1,1}}\Ratio{833}{2279} &
{\cellcolor[rgb]{0.996,1,1}}\np{0}/\np{4} &
{\cellcolor[rgb]{0.996,1,1}}\np{1}/\np{1} &
{\cellcolor[rgb]{0.996,1,1}}\np{0}/\np{1} \\
\rowcolor[rgb]{0.91,0.918,0.918} \texttt{java-faker} &
\np{4}/\np{4} &
\ChartSmall[p]{0}{506} &
\Ratio{107}{506} &
\textcolor{red}{\ding{55}} &
\textcolor{teal}{\ding{51}} &
\np{3}/\np{4} &
\ChartSmall[p]{211}{506} &
\Ratio{107}{295} &
\np{0}/\np{4} &
\np{1}/\np{1} &
\np{0}/\np{1} \\
\texttt{jcabi-github} &
{\cellcolor[rgb]{0.996,1,1}}\np{17}/\np{20} &
{\cellcolor[rgb]{0.996,1,1}}\ChartSmall[p]{9}{3921} &
{\cellcolor[rgb]{0.996,1,1}}\Ratio{312}{3912} &
{\cellcolor[rgb]{0.996,1,1}}\textcolor{red}{\ding{55}} &
{\cellcolor[rgb]{0.996,1,1}}\textcolor{teal}{\ding{51}} &
{\cellcolor[rgb]{0.996,1,1}}\np{16}/\np{17} &
{\cellcolor[rgb]{0.996,1,1}}\ChartSmall[p]{2415}{3912} &
{\cellcolor[rgb]{0.996,1,1}}\Ratio{312}{1497} &
{\cellcolor[rgb]{0.996,1,1}}\np{0}/\np{17} &
{\cellcolor[rgb]{0.996,1,1}}\np{1}/\np{1} &
{\cellcolor[rgb]{0.996,1,1}}\np{0}/\np{1} \\
\rowcolor[rgb]{0.91,0.918,0.918} \texttt{jimfs} &
\np{8}/\np{9} &
\ChartSmall[p]{0}{3560} &
\Ratio{124}{3560} &
\textcolor{teal}{\ding{51}} &
&
\np{6}/\np{8} &
\ChartSmall[p]{1741}{3560} &
\Ratio{124}{1819} &
\np{2}/\np{8} &
-- &
-- \\
\texttt{jooby} &
{\cellcolor[rgb]{0.996,1,1}}\np{20}/\np{22} &
{\cellcolor[rgb]{0.996,1,1}}\ChartSmall[p]{0}{6945} &
{\cellcolor[rgb]{0.996,1,1}}\Ratio{320}{6945} &
{\cellcolor[rgb]{0.996,1,1}}\textcolor{red}{\ding{55}} &
{\cellcolor[rgb]{0.996,1,1}}\textcolor{teal}{\ding{51}} &
{\cellcolor[rgb]{0.996,1,1}}\np{19}/\np{20} &
{\cellcolor[rgb]{0.996,1,1}}\ChartSmall[p]{730}{6945} &
{\cellcolor[rgb]{0.996,1,1}}\Ratio{320}{6215} &
{\cellcolor[rgb]{0.996,1,1}}\np{0}/\np{20} &
{\cellcolor[rgb]{0.996,1,1}}\np{1}/\np{1} &
{\cellcolor[rgb]{0.996,1,1}}\np{0}/\np{1} \\
\rowcolor[rgb]{0.91,0.918,0.918} \texttt{lettuce} &
\np{39}/\np{44} &
\ChartSmall[p]{0}{10364} &
\Ratio{1302}{10364} &
\textcolor{red}{\ding{55}} &
\textcolor{teal}{\ding{51}} &
\np{36}/\np{39} &
\ChartSmall[p]{1865}{10364} &
\Ratio{1302}{8499} &
\np{2}/\np{39} &
\np{0}/\np{1} &
\np{1}/\np{1} \\
\texttt{modelmapper} &
{\cellcolor[rgb]{0.996,1,1}}\np{6}/\np{6} &
{\cellcolor[rgb]{0.996,1,1}}\ChartSmall[p]{0}{2700} &
{\cellcolor[rgb]{0.996,1,1}}\Ratio{210}{2700} &
{\cellcolor[rgb]{0.996,1,1}}\textcolor{red}{\ding{55}} &
{\cellcolor[rgb]{0.996,1,1}}\textcolor{teal}{\ding{51}} &
{\cellcolor[rgb]{0.996,1,1}}\np{4}/\np{6} &
{\cellcolor[rgb]{0.996,1,1}}\ChartSmall[p]{66}{2700} &
{\cellcolor[rgb]{0.996,1,1}}\Ratio{210}{2634} &
{\cellcolor[rgb]{0.996,1,1}}\np{1}/\np{6} &
{\cellcolor[rgb]{0.996,1,1}}\np{0}/\np{1} &
{\cellcolor[rgb]{0.996,1,1}}\np{1}/\np{1} \\
\rowcolor[rgb]{0.91,0.918,0.918} \texttt{mybatis-3} &
\np{8}/\np{8} &
\ChartSmall[p]{0}{1345} &
\Ratio{480}{1345} &
\textcolor{red}{\ding{55}} &
\textcolor{teal}{\ding{51}} &
\np{7}/\np{8} &
\ChartSmall[p]{414}{1345} &
\Ratio{480}{931} &
\np{0}/\np{8} &
\np{0}/\np{1} &
\np{1}/\np{1} \\
\texttt{OpenPDF} &
{\cellcolor[rgb]{0.996,1,1}}\np{12}/\np{35} &
{\cellcolor[rgb]{0.996,1,1}}\ChartSmall[p]{2336}{17512} &
{\cellcolor[rgb]{0.996,1,1}}\Ratio{484}{15176} &
{\cellcolor[rgb]{0.996,1,1}}\textcolor{red}{\ding{55}} &
{\cellcolor[rgb]{0.996,1,1}}\textcolor{teal}{\ding{51}} &
{\cellcolor[rgb]{0.996,1,1}}\np{11}/\np{12} &
{\cellcolor[rgb]{0.996,1,1}}\ChartSmall[p]{9155}{15176} &
{\cellcolor[rgb]{0.996,1,1}}\Ratio{484}{6021} &
{\cellcolor[rgb]{0.996,1,1}}\np{0}/\np{12} &
{\cellcolor[rgb]{0.996,1,1}}\np{1}/\np{1} &
{\cellcolor[rgb]{0.996,1,1}}\np{0}/\np{1} \\
\rowcolor[rgb]{0.91,0.918,0.918} \texttt{pdfbox} &
\np{6}/\np{7} &
\ChartSmall[p]{63}{6836} &
\Ratio{754}{6773} &
\textcolor{teal}{\ding{51}} &
&
\np{6}/\np{6} &
\ChartSmall[p]{5070}{6773} &
\Ratio{754}{1703} &
\np{0}/\np{6} &
-- &
-- \\
\texttt{pf4j} &
{\cellcolor[rgb]{0.996,1,1}}\np{3}/\np{3} &
{\cellcolor[rgb]{0.996,1,1}}\ChartSmall[p]{0}{115} &
{\cellcolor[rgb]{0.996,1,1}}\Ratio{93}{115} &
{\cellcolor[rgb]{0.996,1,1}}\textcolor{teal}{\ding{51}} &
{\cellcolor[rgb]{0.996,1,1}} &
{\cellcolor[rgb]{0.996,1,1}}\np{2}/\np{3} &
{\cellcolor[rgb]{0.996,1,1}}\ChartSmall[p]{10}{115} &
{\cellcolor[rgb]{0.996,1,1}}\Ratio{93}{105} &
{\cellcolor[rgb]{0.996,1,1}}\np{1}/\np{3} &
{\cellcolor[rgb]{0.996,1,1}}-- &
{\cellcolor[rgb]{0.996,1,1}}-- \\
\rowcolor[rgb]{0.91,0.918,0.918} \texttt{poi-tl} &
\np{33}/\np{36} &
\ChartSmall[p]{258}{12143} &
\Ratio{255}{11885} &
\textcolor{red}{\ding{55}} &
\textcolor{teal}{\ding{51}} &
\np{27}/\np{33} &
\ChartSmall[p]{5192}{11885} &
\Ratio{255}{6693} &
\np{5}/\np{33} &
\np{0}/\np{1} &
\np{1}/\np{1} \\
\texttt{Recaf} &
{\cellcolor[rgb]{0.996,1,1}}\np{49}/\np{56} &
{\cellcolor[rgb]{0.996,1,1}}\ChartSmall[p]{518}{10769} &
{\cellcolor[rgb]{0.996,1,1}}\Ratio{538}{10251} &
{\cellcolor[rgb]{0.996,1,1}}\textcolor{teal}{\ding{51}} &
{\cellcolor[rgb]{0.996,1,1}} &
{\cellcolor[rgb]{0.996,1,1}}\np{41}/\np{49} &
{\cellcolor[rgb]{0.996,1,1}}\ChartSmall[p]{2952}{10251} &
{\cellcolor[rgb]{0.996,1,1}}\Ratio{538}{7299} &
{\cellcolor[rgb]{0.996,1,1}}\np{8}/\np{49} &
{\cellcolor[rgb]{0.996,1,1}}-- &
{\cellcolor[rgb]{0.996,1,1}}-- \\
\rowcolor[rgb]{0.91,0.918,0.918} \texttt{RxRelay} &
\np{2}/\np{2} &
\ChartSmall[p]{0}{1758} &
\Ratio{16}{1758} &
\textcolor{red}{\ding{55}} &
\textcolor{teal}{\ding{51}} &
\np{1}/\np{2} &
\ChartSmall[p]{9}{1758} &
\Ratio{16}{1749} &
\np{0}/\np{2} &
\np{0}/\np{1} &
\np{1}/\np{1} \\
\texttt{scribejava} &
{\cellcolor[rgb]{0.996,1,1}}\np{7}/\np{8} &
{\cellcolor[rgb]{0.996,1,1}}\ChartSmall[p]{39}{1278} &
{\cellcolor[rgb]{0.996,1,1}}\Ratio{116}{1239} &
{\cellcolor[rgb]{0.996,1,1}}\textcolor{teal}{\ding{51}} &
{\cellcolor[rgb]{0.996,1,1}} &
{\cellcolor[rgb]{0.996,1,1}}\np{6}/\np{7} &
{\cellcolor[rgb]{0.996,1,1}}\ChartSmall[p]{353}{1239} &
{\cellcolor[rgb]{0.996,1,1}}\Ratio{116}{886} &
{\cellcolor[rgb]{0.996,1,1}}\np{1}/\np{7} &
{\cellcolor[rgb]{0.996,1,1}}-- &
{\cellcolor[rgb]{0.996,1,1}}-- \\
\rowcolor[rgb]{0.91,0.918,0.918} \texttt{tablesaw} &
\np{9}/\np{9} &
\ChartSmall[p]{0}{1447} &
\Ratio{7}{1447} &
\textcolor{teal}{\ding{51}} &
&
\np{7}/\np{9} &
\ChartSmall[p]{379}{1447} &
\Ratio{7}{1068} &
\np{2}/\np{9} &
-- &
-- \\
\texttt{tika} &
{\cellcolor[rgb]{0.996,1,1}}\np{2}/\np{2} &
{\cellcolor[rgb]{0.996,1,1}}\ChartSmall[p]{0}{253} &
{\cellcolor[rgb]{0.996,1,1}}\Ratio{435}{253} &
{\cellcolor[rgb]{0.996,1,1}}\textcolor{teal}{\ding{51}} &
{\cellcolor[rgb]{0.996,1,1}} &
{\cellcolor[rgb]{0.996,1,1}}\np{2}/\np{2} &
{\cellcolor[rgb]{0.996,1,1}}\ChartSmall[p]{187}{253} &
{\cellcolor[rgb]{0.996,1,1}}\Ratio{435}{66} &
{\cellcolor[rgb]{0.996,1,1}}\np{0}/\np{2} &
{\cellcolor[rgb]{0.996,1,1}}-- &
{\cellcolor[rgb]{0.996,1,1}}-- \\
\rowcolor[rgb]{0.91,0.918,0.918} \texttt{undertow} &
\np{5}/\np{5} &
\ChartSmall[p]{0}{742} &
\Ratio{1581}{742} &
\textcolor{teal}{\ding{51}} &
&
\np{5}/\np{5} &
\ChartSmall[p]{224}{742} &
\Ratio{1581}{518} &
\np{0}/\np{5} &
-- &
-- \\
\texttt{woodstox} &
{\cellcolor[rgb]{0.996,1,1}}\np{5}/\np{5} &
{\cellcolor[rgb]{0.996,1,1}}\ChartSmall[p]{0}{864} &
{\cellcolor[rgb]{0.996,1,1}}\Ratio{208}{864} &
{\cellcolor[rgb]{0.996,1,1}}\textcolor{red}{\ding{55}} &
{\cellcolor[rgb]{0.996,1,1}}\textcolor{teal}{\ding{51}} &
{\cellcolor[rgb]{0.996,1,1}}\np{3}/\np{5} &
{\cellcolor[rgb]{0.996,1,1}}\ChartSmall[p]{34}{864} &
{\cellcolor[rgb]{0.996,1,1}}\Ratio{208}{830} &
{\cellcolor[rgb]{0.996,1,1}}\np{0}/\np{5} &
{\cellcolor[rgb]{0.996,1,1}}\np{1}/\np{2} &
{\cellcolor[rgb]{0.996,1,1}}\np{1}/\np{2} \\
\hline
\textsc{Total} &
{\cellcolor[rgb]{0.996,1,1}}\np{396}/\np{467} &
{\cellcolor[rgb]{0.996,1,1}}\ChartSmall[p]{5813}{135343} &
{\cellcolor[rgb]{0.996,1,1}}\Ratio{15594}{129530} &
\multicolumn{1}{r|}{{\cellcolor[rgb]{0.996,1,1}}\np{14}/\np{30}} &
\multicolumn{1}{r|}{{\cellcolor[rgb]{0.996,1,1}}\np{16}/\np{30}} &
{\cellcolor[rgb]{0.996,1,1}}\ChartSmall[p]{343}{396} &
{\cellcolor[rgb]{0.996,1,1}}\ChartSmall[p]{51631}{129530} &
{\cellcolor[rgb]{0.996,1,1}}\Ratio{15594}{77899} &
{\cellcolor[rgb]{0.996,1,1}}\np{32}/\np{396} &
{\cellcolor[rgb]{0.996,1,1}}\np{12}/\np{21} &
{\cellcolor[rgb]{0.996,1,1}}\np{9}/\np{21} \\
\hline
\end{tabular}
\end{table*}

With this first research question, we set a baseline to assess the impact of dependency specialization regarding the reduction of the number of classes in third-party dependencies.
To do so, we report the number of classes removed through state-of-the-art dependency debloating with \depclean, as described in \autoref{sec:protocol-rq1}.
We report the ratio of third-party classes remaining after debloating, with respect to the number of classes in each project presented in \autoref{tab:descriptive}.

For our \np{30} study subjects, the column \nbcd in \autoref{tab:deptrim} denotes the number of \compile-scope dependencies that remain after identifying and removing bloated dependencies with \depclean, over the original number of \compile-scope dependencies in the project (column \#\cd in \autoref{tab:descriptive}).
In total, \depclean removes \nbOfDependenciesDebloated  bloated dependencies, with a median of \np{8} dependencies, across the \np{30} projects.
\depclean removes \np{23} bloated dependencies from \texttt{OpenPDF}, which is the largest number of bloated dependencies for one project in our dataset.
In total, \depclean removes \np{5813}  third-party classes.
It is interesting to note that, for all the projects, dependency debloating removes \perOfClassesRemovedWithDepclean of the total number of classes.



All projects have at least \np{2} NBCDs, while \texttt{Recaf} has the maximum number of NBCDs at \np{41}.
In \np{13} projects, such as \texttt{classgraph} and \texttt{commons-validator}, all the dependencies are used. 
Therefore, executing \depclean does not contribute to the removal of any class on those projects.
On the other hand, we find that in \np{5} projects, the bloated dependencies do not contain \class files at all, such as in the case of \texttt{flink}.
This happens when a bloated dependency only contains assets, such as resource files, or is explicitly designed to avoid conflicts with other dependencies~\cite{wang2021will}.
For example, the dependency \texttt{com.google.guava:listenablefuture} is present in the dependency tree of \np{5} projects, and it is  intentionally empty to avoid conflicts with \texttt{guava}~\cite{listenablefuture}.
Another dependency, called \texttt{batik-shared-resources}, is included in the dependency tree of \np{2} projects, and only contains resource files.
We investigate the nature of these resource files and find that they are dependency license statements and build-related metadata. 
Thus, the removal of such dependencies does not result in a build failure within the projects.

The column \ratiod in \autoref{tab:deptrim} presents the ratio of the number of classes in the NBCDs to the original number of classes in the project (column \textsc{Project} in \autoref{tab:descriptive}).
For \np{11} of the \np{30} projects, \ratiod is less than \ratioo from \autoref{tab:descriptive}.
This corresponds to cases where debloating results in fewer third-party classes in the compiled project.
For example, the removal of the \np{23} bloated dependencies from \texttt{OpenPDF} results in the maximum reduction in the number of classes (\np{2336}). Consequently, \ratiod for \texttt{OpenPDF} is \np{31.4}$\times$, which is \np{4.8} less than its \ratioo.
The project with the highest percentage of classes removed is \texttt{graphhopper}, for which the removal of \np{5} bloated dependencies leads to a \np[\%]{23.9} reduction in the number of third-party classes. \ratiod for \texttt{graphhopper} is \np{6.6}$\times$, down from its original \ratioo of \np{8.7}$\times$.
However, despite debloating dependencies, the total \ratiod across the \np{30} projects is \ratioDebloated, which is only \np{0.4} less than the total original \ratioo.

Of the \np{9} \compile-scope dependencies in \jacop (column \textsc{\#CD} in \autoref{tab:descriptive}), \depclean identifies \np{5} dependencies as bloated and removes them. 
This leads to the removal of \np{504} classes, and \np{4} remaining \textsc{NBCD}s with a total of \np{7983} classes.
Correspondingly, \ratiod reduces to \np{9.6}$\times$, down from the original \ratioo of \np{10.2}$\times$.
The \np{4} NBCDs are the target for specialization with \deptrim, which will remove unused classes within these dependencies while ensuring that \jacop still correctly builds.

\begin{mdframed}[style=mpdframe]
\textbf{Answer to RQ1:} 
State-of-the-art dependency debloating with \depclean contributes to the removal of \nbOfDependenciesDebloated bloated dependencies from \np{30} real-world Java projects. This corresponds to the removal of \perOfClassesDebloated third-party classes in total.
Yet, the dependency classes to project classes ratio is reduced by only \np{0.4} (from \ratioClassesInProjectWrtClassesInDependenciesOriginal to \ratioDebloated). 
This calls for more extensive code removal to reduce the dependencies to the strictly necessary parts. 
\end{mdframed}

\subsection{RQ2: \RQtwo}\label{sec:results-rq2}

This research question evaluates the ability of \deptrim to perform automatic dependency specialization for the study subjects described in \autoref{sec:study-subjects}. 
We consider the specialization procedure to be successful if \deptrim produces a valid set of specialized dependencies, with a corresponding specialized dependency tree captured in a \pom, and for which the project builds correctly. 
To reach this successful state, the project to be specialized must pass through all the build phases of the \maven build lifecycle, \ie, compilation, testing, and packaging, according to the protocol described in~\autoref{sec:protocol-rq1}.\looseness=-1 

Columns \textsc{\tst}, \textsc{\pst}, and \textsc{NBCD Specialized} in \autoref{tab:deptrim} present the results obtained.
First, we observe that for a total of \ratioTstThatPass projects, \deptrim produces a totally specialized tree (\tst), \ie, the project builds successfully with a specialized version of all its non-bloated \compile-scope dependencies.
For these projects, \deptrim successfully identifies and removes unused classes within the dependencies.
Moreover, \deptrim updates the dependency tree of each project by replacing original dependencies with specialized ones. 
\revision{The projects correctly compile, and their original test suite still passes, indicating that their behavior is intact \wrt the tests, despite the dependency tree specialization. }
Overall, these results confirm that dependency specialization is feasible for real-world projects.

We illustrate {\tst}s with the example of \texttt{pdfbox}, a utility library and tool to manipulate PDF documents. \deptrim specializes the \np{6} {\nbcd}s of \texttt{pdfbox}, and builds its \tst successfully. 
Of these \np{6} dependencies, \np{4} are direct: \texttt{fontbox}, \texttt{commons-logging}, \texttt{pdfbox-io}, and \texttt{bcprov-jdk18on}; whereas \np{2} are transitive: \texttt{bcutil-jdk18on} and \texttt{bcpkix-jdk18on}. 
Another interesting example is \texttt{guice}, a popular dependency injection framework from Google branded as a “lightweight” alternative to existing libraries, as stated in its official documentation.
\deptrim builds \texttt{guice} with a \tst, thus making it even smaller.
The project that builds with a \tst and has the largest number of specialized dependencies is \texttt{Recaf} with \np{41} specialized dependencies. 
Note that, when specializing transitive dependencies, \deptrim keeps all the classes in the direct dependencies that are necessary to access the APIs in transitive dependencies, whether directly or indirectly.
For example, the transitive dependency \texttt{commons-lang3} in \texttt{Recaf} is resolved and used from the direct dependency \texttt{jphantom}, a Java library for program complementation~\cite{balatsouras2013class}.
Thus, \deptrim keeps the bytecode in \texttt{commons-lang3} that is necessary to access the used features provided by \texttt{jphantom}.


On the other hand, projects that do not build with a \tst signify cases where at least one \compile-scope dependency relies on dynamic Java features that make static analysis unsound. 
This observation is in line with previous work showing that Java reflection and other dynamic features impose limitations on performing static analysis in the Java ecosystem~\cite{li2019understanding}.
\revision{However, even for these projects, \deptrim successfully builds a partially specialized tree (\pst) by targeting dependencies that could be specialized, and discarding the ones that cause the build to break.}
In total, \ratioPstThatPass of the projects build successfully with a \pst.
In these cases, \deptrim successfully identifies the subset of dependencies that are safe for specialization, and validates that the projects still correctly build with a \pst.


For example, \deptrim successfully specializes \np{4} dependencies in the project \texttt{modelmapper}, an object mapping library that automatically maps objects to each other. \deptrim creates a \pst with which \texttt{modelmapper} builds successfully.
Note that none of the \np{6} \compile-scope dependencies of \texttt{modelmapper} are bloated, and hence debloating the project with \depclean has no impact on it. 
However, after executing \deptrim, the direct dependencies \texttt{objenesis} and \texttt{asm-tree} are specialized.
Moreover, the transitive dependencies \texttt{asm-commons} and \texttt{asm}, resolved from \texttt{asm-tree} are also specialized.
This example illustrates the impact of specialization beyond dependency debloating, for projects that build with a \pst. 
Indeed, across our study subjects, there are \np{8} projects that successfully build with a \tst, and \np{5} that build with a \pst, and yet for which no classes are removed through \depclean.


\begin{table}
\centering
\footnotesize
\caption{Dependencies specialized in the \texttt{OpenPDF} project}
\label{tab:openpdf}
\begin{tabular}{|l|r|r|} 
\hline
\textsc{Dependency}          & \textsc{Type} & \textsc{Classes Removed}      \\ 
\hline
\rowcolor[rgb]{0.91,0.918,0.918} 
\texttt{xmlgraphics-commons} & Transitive    & \ChartSmall[q]{366}{375}      \\
\texttt{bcutil-jdk18on}      & Transitive    & \ChartSmall[q]{532}{579}      \\
\rowcolor[rgb]{0.91,0.918,0.918} 
\texttt{fop-core}            & Transitive    & \ChartSmall[q]{2278}{2547}    \\
\texttt{bcpkix-jdk18on}      & Direct        & \ChartSmall[q]{697}{841}      \\
\rowcolor[rgb]{0.91,0.918,0.918} 
\texttt{bcprov-jdk18on}      & Direct        & \ChartSmall[q]{5696}{7149}    \\
\texttt{xml-apis}            & Transitive    & \ChartSmall[q]{234}{346}      \\
\rowcolor[rgb]{0.91,0.918,0.918} 
\texttt{fontbox}             & Transitive    & \ChartSmall[q]{100}{157}      \\
\texttt{xalan}               & Transitive    & \ChartSmall[q]{942}{1501}     \\
\rowcolor[rgb]{0.91,0.918,0.918} 
\texttt{icu4j}               & Direct        & \ChartSmall[q]{578}{1555}     \\
\texttt{serializer}          & Transitive    & \ChartSmall[q]{69}{108}       \\
\rowcolor[rgb]{0.91,0.918,0.918} 
\texttt{commons-logging}     & Transitive    & \ChartSmall[q]{6}{18}         \\
\texttt{fop}                 & Transitive    & N/A                           \\ 
\hline
\textsc{Total}               & \np[D]{3}/\np[T]{9}             & \ChartSmall[q]{9155}{15176}  \\
\hline
\end{tabular}
\end{table}

It is interesting to notice that some of our study subjects share dependencies that are specialized.
For example, \texttt{slf4j-api} is specialized in \np{12} different projects, \texttt{jsr305} in \np{8} projects, and \texttt{commons-io} in \np{4} projects.
The projects \texttt{jcabi-github}, \texttt{jooby}, and \texttt{Recaf} include all these three dependencies in their dependency tree. 
After investigating the contents of the specialized versions of \texttt{slf4j-api} prepared by \deptrim, we find that there are three sets of variants for which this dependency contains the same number of classes.
Thus, deploying multiple specialized versions of \texttt{slf4j-api} to external repositories can contribute to reducing its attack surface for projects that reuse the exact same features.
This specialized form of code reuse also increases software diversity.
Furthermore, the dependency \texttt{bcprov-jdk18on}, which contains the largest number of classes among the dependencies (\np{3768}), is successfully specialized in \np{2} projects, \texttt{OpenPDF} and \texttt{pdfbox}.
Our findings suggest that specializing dependencies with a large number of classes yields a greater reduction of third-party code.
To confirm this hypothesis, additional investigation is required.

Our experiments show that, despite the challenges of specializing the dependency trees of our \np{30} real-world study subjects, \deptrim is capable of specializing \nbSpecializedDependencies of the \nbNBCD non-bloated \compile-scope dependencies across them.
A key aspect of our evaluation is that we validate that each project builds successfully using its specialized dependency tree.
We manually analyze and classify the cases where specialization is not achievable for a dependency, in RQ4.
The specialized dependencies contribute to the deployment of smaller project binaries, to reduce their attack surface, and to increase dependency diversity when deployed to external repositories.

\begin{mdframed}[style=mpdframe]
\textbf{Answer to RQ2:} 
\deptrim  successfully builds \nbTstThatPass real-world projects with a totally specialized dependency tree. For the other \nbPstThatPass projects, \deptrim finds the largest subset of specialized dependencies that do not break the build. In total, \deptrim specializes \ratioOfSpecializedDependencies of the non-bloated \compile-scope dependencies. This is evidence that a large majority of dependencies in Java projects can be specialized without impacting the project build.
\end{mdframed}

\subsection{RQ3: \RQthree}\label{sec:results-rq3}

To answer our third research question, we count the number of classes removed by \deptrim in the  \nbSpecializedDependencies successfully specialized dependencies. 
The goal is to evaluate the effectiveness of \deptrim in removing unused \class files through specialization, as described in \autoref{sec:protocol-rq2}. 
We also report the impact of this reduction on the ratio of third-party classes to project classes, \ie, \ratios. 


 
The column \textsc{Classes Removed} in \autoref{tab:deptrim} shows the number of classes removed by \deptrim from the \textsc{NBCD}s of each project, in order to build its \tst or \pst. 
\deptrim removes a total of \nbTotalClassesRemoved classes, with a median removal of \np{858} third-party classes for each project.
This represents \percOfClassesRemoved of the total number of classes in the third-party dependencies for all the projects (\ie, \np{135343} per \autoref{tab:descriptive}). 
For example, the project \texttt{tika} has \np{2} dependencies specialized: \texttt{slf4j-api} with \ChartSmall[q]{9}{52} classes removed, and \texttt{commons-io} with \ChartSmall[q]{178}{201} classes removed.
This represents a removal of \ChartSmall[p]{187}{253} third-party classes in \texttt{tika}, as a result of which its \ratios is \Ratio{435}{66}.
Thus, the ratio of dependency classes to project classes in \texttt{tika} decreased by \np{0.4} compared to \ratiod (\ie, \np{0.6}$\times$). 


The project with the highest percentage of dependency classes removed is  \texttt{commons-validator} with \np[\%]{80.1}, \ie, \np{625} of the \np{780} original third-party classes.
This drastic reduction is the result of successfully removing unused classes from all its \compile-scope dependencies: \texttt{commons-beanutils}, \texttt{commons-collections}, \texttt{commons-digester} and \texttt{commons-logging}.
In particular, \deptrim identifies that only \np{7} of the \np{460} classes in \texttt{commons-collections} are required by \texttt{commons-validator}.
These classes support an implementation of the \texttt{HashMap} data structure for multi-threaded operations, which are used by \texttt{commons-validator} for processing form fields.
The other \np{453} types within \texttt{commons-collections} correspond to data structures that are not required by \texttt{commons-validator}, and are consequently removed by \deptrim.

The project with the largest number of classes removed (\np{9155}) is \texttt{OpenPDF}.
\autoref{tab:openpdf} shows the dependencies specialized in \texttt{OpenPDF}, of which \np{3} are direct and \np{9} are transitive.
\deptrim builds \texttt{OpenPDF} with a \pst, excluding the dependency \texttt{fop} from the specialized dependency tree. Looking at the \np{11} successfully specialized dependencies, we observe that \texttt{OpenPDF} depends transitively on a family of dependencies from the Apache XML Graphics Project, including \texttt{fop-core} and \texttt{xmlgraphics-commons}, from which \deptrim removes \ChartSmall[p]{2278}{2547} and \ChartSmall[p]{366}{375} unused classes, respectively.
\texttt{OpenPDF} also depends directly on \texttt{bcpkix-jdk18on} and \texttt{bcprov-jdk18on}, which are dependencies from the Bouncy Castle family of cryptographic libraries, from which \ChartSmall[p]{697}{841} and \ChartSmall[p]{5696}{7149} classes are removed, respectively. 
Moreover, \deptrim systematically identifies functionalities that are used transitively through direct dependencies.
For example, two classes within \texttt{OpenPDF}, called \texttt{PdfPKCS7} and \texttt{TSAClientBouncyCastle}, use classes from the direct dependency \texttt{bcpkix-jdk18on}. In turn, these classes of \texttt{bcpkix-jdk18on} depend on \np{4} classes within \texttt{bcutil-jdk18on} that are responsible for supporting the encoding of the Time Stamp Protocol. 
Therefore, \deptrim marks these \np{4} classes within the transitive dependency as necessary for \texttt{OpenPDF}, and does not remove them. 
Note that \texttt{OpenPDF} built with a specialized dependency tree may be deployed to an external repository, which reduces the attack surface of the clients that rely on the features that are provided by \texttt{OpenPDF} when used as a library.

\begin{table}
\centering
\footnotesize
\caption{\revision{Summary of global impact of \deptrim: number of classes in the compile-scope dependencies (CD) and size of the bytecode of the compile-scope dependencies (CD) in the original projects, and impact of \deptrim in reducing the number of classes and the size of bytecode for dependencies.}}
\label{tab:jarsizes}
\setlength\tabcolsep{3pt}
\begin{tabular}{|l|r|r|r|r|} 
\hline
& \textsc{Min.} & \textsc{Max.} & \textsc{Med.}   & \textsc{Total} \\ 
\hline
\textsc{Original}  &  &  &  &  \\
\hline
\rowcolor[rgb]{0.91,0.918,0.918} 
\# \textsc{CD} classes   & \np{115} & \np{17512} & \np{3130} & \np{135343} \\
CD bytecode size (MB) & \np{0.4} & \np{49.5} & \np{9.7} & \np{455.9} \\
\hline
\textsc{Impact of \deptrim}  &  &  &  &  \\
\hline
\rowcolor[rgb]{0.91,0.918,0.918} 
\# \textsc{CD} classes removed & \np{9} & \np{11491} & \np{873.5} & \np{57444} (\np[\%]{42.2})\\
CD bytecode reduction (MB) & \np{0.02} & \np{37.1} & \np{1.9} & \np{186.1} (\np[\%]{35.7})\\

\hline

\end{tabular}
\end{table}

\revision{
Overall, the specialization of non-bloated dependencies can significantly reduce the share of third-party classes, beyond state-of-the-art dependency debloating techniques such as \depclean~\cite{soto2021comprehensive}.
This is evidenced in the last row of \autoref{tab:deptrim}, where we report the removal of \np{51631} classes, just through specialization, which represents \np[\%]{39,9} of the classes in non-bloated dependencies. While these observations are compelling evidence of the benefits of specialization, we now reflect upon the overall effect of \deptrim.}

\revision{
\autoref{tab:jarsizes} summarizes the key metrics about the impact of \deptrim on dependency trees. 
First, we provide the distribution of the number of classes in compile-scope dependencies (CD), as well as the distribution of the size of the bytecode for these dependencies.
Second, we provide the global reduction  of the number of classes and the size of the bytecode for third-party dependencies. 
While we focused on the number of classes so far, here we also include the impact on the size of the bytecode, as it is an important performance metric for some applications. 
Specifically, the table highlights the minimum, maximum, median, and total values of CD classes and bytecode sizes. For example, 
\deptrim removes \np{11491} CD classes in total for \texttt{OpenPDF}, which is the maximum number of CD classes removed.
Per \autoref{tab:deptrim}, we see that \np{2336} classes were removed through dependency debloating, while \np{9155} classes were removed the specialization of \np{11} dependencies.
In \texttt{RxRelay}, \deptrim removes only 9 classes, exclusively by specializing one dependency.
In total, \deptrim succeeds in removing \np[\%]{42.2} of the CD classes.
This is a significant reduction in the prevalence of third-party code in the Java projects under study.
For \jacop, \deptrim achieves the largest reduction in the size of third-party bytecode, removing 37.1 MB.
This is essentially due to large dependencies in \jacop such as \texttt{scala-compiler}, for which \deptrim removes \np{2982} classes (see \autoref{fig:jacop-dt-specialized}).
Note that \deptrim removes more than 1.9 MB of third-party bytecode for half the study subjects.
In total, \deptrim removes 186.1 MB of third-party bytecode, which corresponds to a reduction in \np[\%]{35.7}.
}

\begin{mdframed}[style=mpdframe]
\textbf{Answer to RQ3:} 
\deptrim reduces the number of classes in the  dependency tree of each of the \np{30} projects.
\revision{Overall, by adding bytecode specialization to dependency debloating, \deptrim reduces dependency classes by a total of \np[\%]{42.2} and the size of dependency bytecode by \np[\%]{35.7}.}
The dependency classes to project classes ratio reduces from \ratioClassesInProjectWrtClassesInDependenciesOriginal  in the original project to \ratioDeptrim.
Dependency specialization drastically reduces the share of third-party bytecode in Java projects.
\end{mdframed}

\subsection{RQ4: \RQfour}\label{sec:results-rq4}

With this research question we report on the cases where there is no scope for specialization in a non-bloated dependency, as well as cases where projects do not build successfully with a specialized dependency in the dependency tree.

First, we observe that \np{14} projects include at least one dependency that is totally used.
A total of \nbTUD dependencies are totally used by their respective client projects, as presented in the column \textsc{TUD} in \autoref{tab:deptrim}.
A dependency is a TUD for a project if all its \texttt{class} files are exercised by the project. 
Consequently, there is no scope for the specialization of a TUD.
TUDs represent \PercentageOfTUD of the non-bloated dependencies.
\texttt{Recaf} has \np{8} TUDs, the largest number in the study subjects. 
Note that a project with TUDs in its dependency tree can still successfully build with a \tst, as is true for \np{8} projects, including \texttt{Recaf}.
We observe that the dependencies \texttt{asm-tree}, \texttt{failureaccess}, and \texttt{minimal-json} are TUDs in \np{2} projects.
For example, \texttt{minimal-json} is totally used by both \texttt{tablesaw} and by \texttt{Recaf}, which is evidence of a minimal API that is completely used by these projects.
As far as we know, this is the first time in the literature that totally used dependencies are identified and quantified.

\deptrim builds \np{16} projects with a \pst.
For example, \deptrim marks \np{1} of the \np{4} non-bloated \compile-scope dependencies in \texttt{java-faker} as not suitable for specialization. This is because building \texttt{java-faker} with the specialized version of \texttt{org.yaml:snakeyaml} prevents the compilation of the project, which includes the \texttt{Lebowski} class. Consequently, \texttt{org.yaml:snakeyaml} is excluded from the specialized dependency tree of \texttt{java-faker} and \deptrim outputs a partially specialized tree that  successfully builds \texttt{java-faker} and includes three specialized dependencies.


\begin{table}
\centering
\footnotesize
\caption{Number of unique failing tests, and the specialized dependency that causes these failures, in the \np{9} projects with tests failures.}
\label{tab:tests}
\begin{tabular}{|l|r|r|} 
\hline
\textsc{Project}       & \textsc{Dependency}      & \textsc{\# Test Fail}                         \\ 
\hline
\rowcolor[rgb]{0.91,0.918,0.918} 
\texttt{checkstyle}    & \texttt{Saxon-HE}        & \ChartSmall[q]{1}{3887}                       \\
\texttt{Chronicle-Map} & \texttt{chronicle-wire}  & \ChartSmall[q]{3}{1231}                       \\
\rowcolor[rgb]{0.91,0.918,0.918} 
\texttt{flink}         & \texttt{commons-math3}   & \ChartSmall[q]{1}{820}                        \\
\texttt{lettuce-core} & \texttt{micrometer-core} & \ChartSmall[q]{6}{2600}                       \\
\rowcolor[rgb]{0.91,0.918,0.918} 
\texttt{modelmapper}   & \texttt{byte-buddy-dep}  & \ChartSmall[q]{4}{618}                        \\
\texttt{mybatis-3}     & \texttt{slf4j-api}       & \ChartSmall[q]{1}{1699}                       \\
\rowcolor[rgb]{0.91,0.918,0.918} 
\texttt{poi-tl}        & \texttt{commons-io}     & \multicolumn{1}{r|}{\ChartSmall[q]{107}{125}}  \\
\texttt{RxRelay}       & \texttt{rxjava}          & \ChartSmall[q]{6}{64}                         \\
\rowcolor[rgb]{0.91,0.918,0.918} 
\texttt{woodstox}      & \texttt{msv-core}        & \ChartSmall[q]{1}{868}                        \\ 
\hline
\textsc{Total}         & \np{9}  & \ChartSmall[q]{130}{11912}                    \\
\hline
\end{tabular}
\end{table}

Column \textsc{Comp. Error} in \autoref{tab:deptrim} shows the number of specialized dependencies for which the build fails due to compilation errors.
This occurs for \nbPstWithCompilationErrors projects and \np{12} dependencies.
We investigate the causes of compilation errors by manually analyzing the logs of the \texttt{maven-compiler-plugin}. We find the following \np{4} causes for compilation to fail:

\begin{itemize}
    \item Some classes are not found during compilation. For example,  attempting to build  \texttt{checkstyle} with specialized versions of  \texttt{commons-beanutils} and \texttt{guava} fails due to the missing classes, \texttt{BasicDynaBean} and \texttt{ClassPath}. Both classes enable dynamic scanning and loading of classes at runtime.
    \item The project has a plugin that fails at compile time. For example, the plugin \texttt{snakeyaml-codegen-maven-plugin} in the project \texttt{helidon} adds code to the project's compiled sources automatically~\cite{SnakeYAML}, and fails when building with the specialized dependency \texttt{smallrye-open-api-core} because the specialization process changes the expected dependency bytecode.
    \item The project has a plugin that checks the integrity of the specialized dependency. For example, the dependency \texttt{commons-io} in project \texttt{jcabi-github} uses the \texttt{maven-enforcer-plugin} to check for certain constraints, including checksums, on the dependency bytecode.
    \item The specialized dependency is not found in the local repository. For example, the specialized dependency \texttt{snakeyaml} in project \texttt{java-faker} is not deployed correctly due to a known issue in this dependency when using the \texttt{android} \maven tag classifier~\cite{JavaFakerIssue}.
\end{itemize}

We now discuss the number of specialized dependencies for which the build reports test failures (column \textsc{Test Fail.} in \autoref{tab:deptrim}).
For \nbPstWithTestFailures projects, one specialized dependency has at least one test failure.
\revision{\deptrim preserves the original tested behavior (\ie, all the tests pass) of \ratioPstTestPass specialized, non-bloated compile-scope dependencies.}
\revision{This high rate of test success is a fundamental result to ensure that the specialized version of the dependency tree preserves the tested behavior of the project.}

In total, we execute \nbTotalTestsExecuted unique tests across all projects (per \autoref{tab:descriptive}). Of these, \nbTotalTestsExecutedThatFail do not pass. 
\deptrim  produces specialized dependency trees that break a few test cases.
These cases reveal the challenges of dependency specialization concerning static analysis.
For example, \deptrim can miss some used classes, resulting in the removal of bytecode that is necessary at runtime.
This is a general constraint for static analysis tools when processing Java applications that rely on dynamic features to load and execute code at runtime.
As a result of the absence of bytecode from a specialized dependency, \np{9} projects report test failures, \eg, an unreachable class loaded at runtime causing a failing test that stops the execution of the build.

\autoref{tab:tests} shows the number of unique test failures (column \textsc{\#Test Fail.}) in the \nbPstWithTestFailures projects that have at least one \pst with test failures, as well as the specialized dependency that causes the failure (column \textsc{Dependency}).
For example, the project \texttt{Chronicle-Map} has \np{3} tests that fail when specializing the dependency \texttt{chronicle-wire}, from a total of \np{1231} executed tests, which represents \ShowPercentage{3}{1231} of the total.
The project with the largest number of test failures is \texttt{poi-tl}, with \ChartSmall[p]{107}{125} tests failures when specializing its dependency \texttt{commons-io}.
Overall, the number of test failures accounts for \PercentageOfTestsExecutedThatFail of the total tests executed in the \np{9} projects, and only \np[\%]{0.5} across the \np{30} projects.

We further investigate the causes of the failures.
To do so, we manually analyze the logs of the tests, as reported by the \texttt{maven-surefire-plugin}. 
We find the following \np{3} causes:

\begin{itemize}
    \item The tests load dependency classes dynamically. For example,  \texttt{poi-tl} relies on the method \texttt{byte[]} in class \texttt{IOUtils} of \texttt{commons-io} to check the size of a file. This method is loaded via reflection through an external configuration file and causes the failure of \np{107} tests.
    \item Some tests  rely on Java serialization  to manipulate objects at runtime, and the input stream  is not closed properly because \deptrim removes a class responsible for closing the input stream. For example, the project \texttt{Chronicle-Map} uses the dependency \texttt{chronicle-wire} for serialization, and  \np{3} tests fail due to a \texttt{ClosedIORuntimeException}.
    \item The project has tests that rely on dependencies that use Java Native Interfaces (JNI) to execute machine code at runtime. For example, the test \texttt{TestWsdlValidation} in project \texttt{woodstox} relies on dependency \texttt{msv-core} which uses JNI to validate XML schemas. \deptrim's static analysis is limited to Java bytecode, and therefore native code executed in third-party dependencies is not considered as used when building the call graph.
\end{itemize}

Our results reveal the challenges of dependency specialization based on static analysis (see \autoref{sec:call-graph}) for real-world Java projects. 
Handling these cases to achieve \np[\%]{100} correctness requires specific domain knowledge of the project, and of the reachable code in the dependencies that exercise some form of dynamic Java features.
To facilitate this task, we provide a dedicated parameter \texttt{ignoreDependencies} in \deptrim so that developers can declare a list of dependency coordinates to be ignored by \deptrim during the call graph analysis.
Nevertheless, we recommend always checking that the build passes to avoid semantic errors when performing bytecode removal transformations.

\begin{mdframed}[style=mpdframe]
\textbf{Answer to RQ4:} 
Of the 396 dependencies that are targets for specialization, \nbTUD are not specialized because they are totally used, 
\ratioPstCompilationErrors dependencies cause a compilation failure when specialized, and \ratioPstTestFailures  lead to a failure at runtime.
For the latter, the test failures represent only \np[\%]{0.5} of the total number of tests executed.
This behavioral assessment of \deptrim demonstrates that the specialized dependency trees preserve a large majority of syntactic and semantic correctness for the \np{30} projects.
\end{mdframed}

\section{Discussion}

In this section, we discuss the state-of-the-art and the current challenges of code specialization in Java, as well as the implications of specialization for software integrity. We also discuss the threats to the validity of our results.

\subsection{Specialization in the Modern Java Ecosystem}\label{sec:spec}

The Java community is currently making substantial efforts to reduce the amount of bloated code  deployed in production.
The GraalVM native image compiler~\cite{ProjectLeyden2022} is perceived by many as an important step in this direction. 
GraalVM relies on static analysis to build a native executable image that only includes the elements reachable from an application entry point and its third-party dependencies~\cite{GraalVMNativeImage2023}. 
To do so, GraalVM operates under a closed-world assumption~\cite{Wimmer2019}: 
all the bytecode  that can be called at runtime, must be known at build time, \ie, when the native-image tool in GraalVM is building the standalone executable ~\cite{wimmer2021graalvm}.
Thanks to this condition, GraalVM is able to perform a set of aggressive optimizations such as the elimination of unused code from third-party dependencies. 
The self-contained native executable image  includes only code that is actually necessary to build and execute a Java project. 
This reduces the size of container images, making Java applications easy to ship and deploy directly in a containerized environment, as microservices for example.


A big challenge is that many legacy applications are not designed according to the closed-world assumption. In this case, the reachability of some bytecode elements (such as classes, methods, or fields) may not be identified due to the Java dynamic features, \eg, reflection, resource access, dynamic proxies, and serialization~\cite{sui2018soundness, landman2017challenges,li2019understanding, reif2019judge}.
For example, the popular dependency \texttt{netty}, an asynchronous event-driven framework, heavily relies on dynamic Java features to perform blocking and non-blocking sockets between servers and clients.
The closed-world constraint of GraalVM imposes strict limits on the natural dynamism of Java upon which libraries and frameworks like \texttt{netty} depend. 
There is a risk of violating the close world assumption if at least one of the dependency in a project relies on some dynamic Java feature. 

To bridge the gap between the requirements of GraalVM and the current state of Java systems, the community is creating new versions of libraries that adhere to the closed-world assumption. 
\revision{In the long run, Java developers will have the option to embrace the full closed-world constraint in order to produce fully-static images. 
Between now and then, however, the community works on developing and delivering incremental improvements which developers can use sooner rather than later.}
\deptrim contributes to this effort, offering a specialization solution for projects that have dependencies potentially conflicting with the closed-world assumption. 
With the creation of a partially specialized tree (\pst), \deptrim effectively achieves dependency specialization without jeopardizing the success of the build, making it a practical option.
\revision{Note that the successful build of a project does not guarantee that its behaviour is unchanged.
Indeed, test suites can never prove the absence of bugs, and must generally concentrate on specific issues, since it is impossible to test everything.
More research is needed to precisely ascertain the extent to which the usage dynamic of features affects dependency specialization.}

\subsection{\revision{Specialized Projects in Production}}\label{sec:specialized-projects-production}

\revision{
In this work, we assess the validity of the specialized projects with respect to their tests suite.
In practice, developers who wish to deploy their specialized project in production, might consider one more validation step to assess how specialization may impact their users.
This additional step depends on how the specialized project is used:  either declared as dependency within client projects, or  deployed as an application with which end-users interact.
}

\revision{
If the specialized project is mainly used as a library, an additional validation step consists in assessing the specialized \jar with respect to representative client projects. 
By running the test suite of these clients, the developers can verify that the specialization does not result in unexpected behaviors.
This mechanism, termed \emph{reverse dependency compatibility testing}, has been applied previously in the literature to identify breaking updates in libraries~\cite{ochoa2022breakbot}.
This requires curating a list of relevant client projects, which usage of the library is meaningful and which test suites actually exercise the library.
If the specialized project essentially faces end users, through a graphical or command-line interface, an additional validation step consists in running a production-like workload on the specialized application. 
}

\revision{
We have performed a proof of concept of this augmented validation for specialized projects to be used in production. We focused on user facing projects. 
Three projects in our dataset have a graphical or command-line interface and can be executed with a workload: \texttt{graphhopper}, \texttt{pdfbox}, and \texttt{Recaf}.
For each of these projects, we build the original \jar and run it with a workload that is representative of a typical production operation.
Next, we build a specialized version of the project with \deptrim, and run the new \jar with the same workload.
The analysis of both executions let's us determine if the observable behavior of the specialized project is as intended.
}

\revision{
The first project, \texttt{graphhopper},\footnote{\url{https://www.graphhopper.com/}} is a routing application based on \texttt{OpenStreetMap}.
\deptrim removes \np{1666} third-party classes from \np{12} dependencies in \texttt{graphhopper} to output a \pst.
The workload for \texttt{graphhopper} consists of running its \jar and fetching the route between four locations in Sweden from its web page.
The second project, \texttt{pdfbox},\footnote{\url{https://pdfbox.apache.org/}} is developed by the Apache Software Foundation.
It offers command-line tools for performing common operations on PDF documents.
\deptrim produces a \tst for \texttt{pdfbox} by specializing \np{6} dependencies within it, which results in the removal of \np{5070} classes.
As the workload for \texttt{pdfbox}, we use $10$ of its command-line utilities, on $5$ PDF documents sourced from \cite{garfinkel2009bringing}.
These operations include text and image extraction, encryption, decryption, and merging and splitting PDF documents.
\texttt{Recaf}\footnote{\url{https://www.coley.software/Recaf/}} is a code editor that allows developers to manipulate bytecode through a graphical user interface. 
\deptrim produces a \tst for \texttt{Recaf}, by removing \np{2952} classes from \np{41} of its dependencies.
As the production workload for \texttt{Recaf}, we import a compiled \texttt{.class} file into its editor, which decompiles it, and renders its source.
We then modify the source by adding a statement, and export this new version as a \texttt{.java} file.
}

\revision{
For all three specialized projects, we do not observe any deviation between the behavior of the original and the specialized version.
The route returned by specialized \texttt{graphhopper} is identical to the one returned by the original.
The \texttt{pdfbox} operations also result in the same output files.
We successfully modify a decompiled \texttt{.class} file with specialized \texttt{Recaf} and export it, as with the original \texttt{Recaf} \jar file.
Additionally, we do not get any unexpected log outputs within \texttt{graphhopper} and \texttt{pdfbox}.
However, \texttt{Recaf} outputs a log message during startup about a missing class within the specialized dependency \texttt{logback-core}.
This non-critical exception can be remediated by adding \texttt{logback-core} to the specialization blacklist of \deptrim.}

The executions of the three projects under realistic workloads confirm that their high-level features are not impacted by specialization.
Developers can leverage dependency specialization to deliver focused versions of their application to end-users, while keeping its behavior intact. 
An interesting direction for future work is to conduct this evaluation for a larger set of specialized projects.

When we run \deptrim across the three case studies, the execution times are in the order of minutes, taking $18$, $10$, and $33$ minutes for \texttt{graphhopper}, \texttt{pdfbox}, and \texttt{Recaf} respectively. 
Execution time variations are due to several factors.
These factors include the test execution time, the number of dependencies a project has, and the extent to which these dependencies are utilized within the project. 
Interestingly, within the \deptrim execution pipeline, the most time-consuming step is not the static analysis. 
Instead, it is the dynamic analysis phase where tests are run with each specialized dependency tree version to validate their usability. 
Given its relatively expedient execution time and the potential benefits of optimizing dependency trees, we advocate for the integration of \deptrim into the Continuous Integration (CI) pipeline of projects, especially during new releases. 
This is particularly beneficial for projects aiming to deploy smaller \jar files for their dependencies, such as those pushing updates to Maven Central, where minimized \jar files could benefit their end users.




\subsection{Specialization and Software Integrity}

The integrity of software supply chains is a timely research topic~\cite{lamb2021reproducible,ahmadvand2019taxonomy,enck2022top}.
Ensuring the integrity of dependencies involves checking that their code has not been tampered with between the moment they are fetched from a repository and the moment they are packaged in the project.
Checksums, such as the SHA family of cryptographic functions, are commonly used to verify the integrity of software dependencies.
These checksums are then integrated as part of the project's software bill of materials (SBOM) that lists all the components that compose it, including open-source libraries, frameworks, and tools~\cite{balliu2023challenges}. 
A comprehensive, well-maintained SBOM can help ensure software integrity by enabling organizations to identify and track potential security vulnerabilities in their software components and take appropriate action to address them, while also complying with regulations and standards. 

The specialization of third-party dependencies modifies the bytecode of the target dependencies, which can break the integrity-checking process.
\revision{In other words, rehashing a dependency with a different bytecode will produce a different hash value, breaking the integrity check.
This is because the checksum of the original bytecode, which was used to verify the integrity of the dependency, will no longer match the checksum of the changed bytecode.}
For example,  \autoref{lst:sha-original} shows a JSON file reporting the checksum of the original dependency \texttt{commons-io} in one of our study subjects, \texttt{jcabi-github}, when using the SHA-256 hashing algorithm. 
\deptrim specializes \texttt{commons-io} by removing unused classes, which constitutes a change in its bytecode, and hence in its checksum, as presented in \autoref{lst:sha-specialized}. 
Therefore, the checksum of the changed bytecode after specialization no longer matches the expected checksum, and the integrity checks fail, as discussed in \autoref{sec:protocol-rq4}. 

A way to ensure the integrity of specialized dependencies is by deploying them to external repositories at build time.
For example, in the previous example, the project \texttt{jcabi-github} could deploy the specialized variant of \texttt{commons-io} to Maven Central with a custom \maven \texttt{groupId}, while updating the checksum in its SBOM accordingly.
This way, it could check the integrity of this dependency against the SHA of the specialized variant. 
This approach provides the benefits of specialization while preserving software integrity.
As far as we know, there is currently no tool that implements this technique.
Preserving integrity in the light of specialization is a challenge for hardening the software supply chain.

\subsection{Threats to Validity}\label{sec:threats-to-validity}

\emph{\textbf{Internal validity.}}
The first internal threat  relates to the usage of static analysis to determine which parts of the dependency bytecode are reachable from the project.
We mitigate this threat, by relying on \depclean, the state-of-the-art tool for debloating Java dependencies \cite{soto2021comprehensive}.
\revision{Another threat lies in  the thoroughness of the test suite, which may not capture all the dependency API behaviors that can be exercised by the project. 
This means that there is a risk that some necessary classes would be removed, yet the build would be successful because of insufficient testing.
For example, the projects \texttt{immutables}, \texttt{scribejava}, and \texttt{tablesaw} successfully build with a \tst but have less than \np{100} tests each.}
To mitigate this threat, we curate a set of study subjects that are mature and contain tests (see \autoref{tab:descriptive}).
\deptrim is a \maven plugin that modifies the \pom on-the-fly during the build process.
It might introduce conflicts between plugins, causing the build to fail. 
For example,  \texttt{maven-enforcer-plugin} or \texttt{license-maven-plugin}  check the \pom to ensure that it meets specific requirements and follows the best practices.
However, since our approach only modifies the code within the entry \texttt{dependencies} in the \pom, the failures due to misconfigurations are minimized. 

\emph{\textbf{External validity.}}
Our results are  representative of the Java ecosystem, and our findings are valid for software projects with these particular characteristics. Moreover, our bytecode removal results are influenced by the number of dependencies of these projects.
To address this, we found our evaluation on \np{30} real-world, well-known projects, derived from sound data sources, as described in \autoref{sec:study-subjects}.
Furthermore, the selected projects cover a variety of application domains (\eg, dependency injection, database handling, machine learning, encryption, IO utilities, \textcolor{violet}{\href{https://youtu.be/dQw4w9WgXcQ}{faking}}, meta-programming, networking, etc).
To the best of our knowledge, this is the largest set of study subjects used in software specialization experiments.

\emph{\textbf{Construct validity.}} 
The threats to construct validity relate to the accuracy and soundness of the results.
Our results may not be reproducible if the projects are compiled with a different Java version or have flaky tests.
To mitigate this threat, we choose the latest Java version and build the original projects two times in order to avoid including projects with flaky tests.
Furthermore, for all RQs, we include logs and automated analysis scripts in our replication package for reproducibility as described in \autoref{sec:evaluation-framework}.

\begin{lstlisting}[language=Pom, float, belowskip=-5 \baselineskip, numbers=none, basicstyle=\ttfamily\footnotesize,frame=trBL,  caption={SHA checksum of the original dependency \texttt{commons-io} in the project \texttt{jcabi-github}}, 
label={lst:sha-original}]
{
  "groupId": "commons-io",
  "artifactId": "commons-io",
  "version": "2.11.0",
  "checksumAlgorithm": "SHA-256",
  "checksum": "961b2f6d87dbacc5d54abf45ab7a6e2495f89b755989 62d8c723cea9bc210908"
}
\end{lstlisting}


\begin{lstlisting}[language=Pom, float, belowskip=-6 \baselineskip, numbers=none, basicstyle=\ttfamily\footnotesize, frame=trBL, caption={SHA checksum of the specialized dependency \texttt{commons-io} in the project \texttt{jcabi-github}}, 
label={lst:sha-specialized}]
{
  "groupId": "se.kth.castor.deptrim.spl",
  "artifactId": "commons-io",
  "version": "2.11.0",
  "checksumAlgorithm": "SHA-256",
  "checksum": "c84eaef6b629729c71a70a2513584e7ccacf70cb4df1  3e38b731bb6193c60e73"
}
\end{lstlisting}

\section{Related Work}\label{sec:related-work}
In this section, we position the contribution of our dependency specialization technique with respect to previous work that aims at reducing the size of applications composed of multiple third-party dependencies.

Several previous works focus on reducing the size of Java applications. While all techniques perform code analysis based on the construction of a call graph, they vary in the way they look for code that can be removed: dead-code removal, inlining, and class hierarchy removal \cite{tip1999practical}; identification and removal of unused optional concerns with respect to a specific installation context \cite{bhattacharya2013combining}, unbundling user-facing application features \cite{FilhoAB16} or tailoring the Java standard library \cite{jiang2016jred,rayside2002extracting}.
In contrast to these efforts that aim at reducing the size of a packaged application, \deptrim targets reduction while keeping the modular structure of the project and its third-party dependencies. Our technique focuses on reducing each dependency while keeping an explicit dependency tree in the form of a specialized \pom file as well as maintaining specialized dependencies as distinct, deployable \jar files.

Bruce~\etal~\cite{bruce2020jshrink} propose \jshrink, augmenting static reachability analysis with dynamic reachability analysis. They rely on test cases to find dynamic features, including methods and fields, invoked at runtime, adding them back to amend the imprecision of static call graphs.
\deptrim differs from \jshrink, as it does not aim to refine reachability analysis to create smaller \jar files of the target project. 
Instead, \deptrim focuses on specializing the dependency tree of a Java project by removing unused code in third-party dependencies independently, such that each dependency can be deployed to external repositories.

In our previous work,  we proposed \depclean, a tool that identifies and removes unused dependencies in the dependency tree~\cite{SotoValero2019,soto2021longitudinal}. 
\depclean constructs a call graph of the bytecode class members by capturing annotations, fields, and methods, and accounts for a limited number of dynamic features such as class literals. 
\depclean produces a variant of the dependency tree without bloated dependencies. 
\deptrim pushes forward the field of dependency debloating through the removal of unused bytecode from individual dependencies, thereby yielding smaller packaged artifacts.

\revision{
\autoref{tab:comparison} shows a comparison between \deptrim and state-of-the-art debloating tools for Java applications, per the recent study of Ponta~\etal~\cite{ponta2021used}.
We have included a recent tool, \textsc{JDBL}, which produces debloated \jar files based on the usage analysis obtained from code coverage tools~\cite{soto2023coverage}.
We compare the tools regarding three distinct specialization outcomes: the specialized project \jar, specialized dependency \jar files, and specialized \pom files. 
A key observation is that state-of-the-art tools primarily focus on generating a debloated \jar file for the target Java project. More specifically, the \maven Shade Plugin, Proguard, and \textsc{JDBL} aim to build an uber-\jar, which encapsulates all the utilized code from the project's dependencies, resulting in a self-contained \jar file. Only \depclean produces a modified version of the \pom file. The key novelty of \deptrim is that it is the first tool that specializes individual dependency \jar files and produces specialized \pom files. The generation and packaging of specialized individual dependency \jar files allows developers to benefit from specialization while maintaining a modular architecture, and they can eventually include the specialized dependencies as part of their project's software bill of materials~\cite{balliu2023challenges}.
}

\begin{table}
\centering
\footnotesize
\caption{\revision{Comparison between specialization outputs of existing Java debloating tools and \deptrim.}}
\label{tab:comparison}
\begin{tabular}{|lccc|}
\hline
\multicolumn{1}{|l}{\multirow{2}{*}{\textsc{Tool}}} & \multicolumn{3}{c|}{\textsc{Specialization Output}}                                      \\ \cline{2-4} 
\multicolumn{1}{|c}{}                               & Project 
\jar & Dependency \jar    & \texttt{POM} file          \\ \hline
Maven Shade                                        & \textcolor{teal}{\ding{51}} &  &                             \\
ProGuard                                           & \textcolor{teal}{\ding{51}} &                             &                             \\
\textsc{JDBL}                                          & \textcolor{teal}{\ding{51}} &                             & \\
\depclean                                          & \textcolor{teal}{\ding{51}} &                             & \textcolor{teal}{\ding{51}} \\
\hline
\deptrim                                           & \textcolor{teal}{\ding{51}} & \textcolor{teal}{\ding{51}} & \textcolor{teal}{\ding{51}} \\ \hline
\end{tabular}
\end{table}
Closely related to \deptrim is the work on code specialization. 
Mishra and Polychronakis propose \shredder~\cite{mishra2018shredder}, a defense-in-depth exploit mitigation tool that protects closed-source applications against code reuse attacks. 
They also build  \saffire~~\cite{mishra2020saffire} which creates specialized and hardened replicas of critical functions with restricted interfaces to prevent code reuse attacks.
These tools target C++ API implementations. They eliminate arguments with static values and restrict the acceptable values of arguments.
A key feature of these techniques is to replace the code of API members by a stub function so that, at runtime, only specialized versions of critical API functions are exposed, while any invocation that violates the enforced policy is blocked.
Focusing on JavaScript applications, Turcotte~\etal~\cite{turcotte2022stubbifier} propose \stubbifier, which replaces unreachable code, identified through static and dynamic call graphs.
\deptrim does not remove unused code from the project but rather replaces dependencies that are partially used by the project with smaller and specialized  versions.

Previous specialization techniques mitigate the risk of removing code that might be needed for a specific execution, by replacing this code by small stub functions. With \deptrim we address the challenges of dynamic language features with another strategy. We specialize each dependency and then assess whether the completely specialized dependency tree still passes the build. If it does not, we search for a partially specialized tree that does not include the dependencies that rely on the dynamic features of Java.
To the best of our knowledge, prior research on software specialization has not addressed the customization of third-party dependencies or the provision of build configuration files to enable the construction of specialized dependency trees. This represents a novel contribution of our work, differentiating it from previous studies in this area.

As part of our experiments with \deptrim, we contribute novel observations to the body of knowledge about library and API usage. Recent work in this area includes the following studies. 
Huang \etal~\cite{huang2022characterizing} study the usage intensity from Java projects to libraries. They find that the number of libraries adopted by a project is correlated to the project size. However, their study does not provide a more fine-grained analysis of the used components.
Hejderup \etal~\cite{hejderup2022prazi} investigate the extent to which Rust projects use the third-party packages in their dependency tree. 
They propose \textsc{Präzi}, a call-based dependency network for \textsc{Crates.io} that operates at the function level.



Some studies examine the benefits of debloating from a security standpoint. 
For instance, Azad \etal~\cite{azad2019less} report that debloating significantly reduces the number of vulnerabilities in web applications, while also making it more difficult for attackers to exploit the remaining ones. Agadakos~\etal~\cite{agadakos2019nibbler} propose \nibbler to erase unused functions within the binaries of shared libraries at the binary level. This enhances existing software defenses, such as continuous code re-randomization and control-flow integrity, without incurring additional run-time overhead. 
\revision{Ye \etal~\cite{ye2021jslim} implement a tool that uses NLP and function call graphs to identify and isolate vulnerabilities in NPM packages, effectively reducing software bloat and preventing known vulnerability exploitation in JavaScript applications.}
Although \deptrim's primary function is to specialize dependency trees and enhance their reusability, it is important to note that the removal of third-party code can lead to a reduction in the potential attack surface.

\section{Conclusion}\label{sec:conclusion}
In this paper, we propose \deptrim, a fully automated technique to specialize third-party dependencies of a Java project. \deptrim systematically identifies and removes unused classes within each reachable dependency, repackages the used classes into a specialized dependency, and replaces the original dependency tree of a project with a specialized version. \deptrim builds a minimal project binary, which only contains code that is necessary for the project.

Our evaluation with \np{30} \maven Java projects demonstrates the capabilities of \deptrim to produce minimal versions of the dependencies in these projects while keeping the original build successful. 
In particular, \deptrim builds totally specialized trees for \nbTstThatPass projects and builds the other \nbPstThatPass with the largest number of specialized dependencies such that the project still builds. 
The ratio of dependency classes to project classes decreases from \ratioClassesInProjectWrtClassesInDependenciesOriginal in the original projects to \ratioDeptrim in the specialized projects. \revision{This represents a reduction of  \np[\%]{35.7} of the  bytecode  size in third-party dependencies. } 
Dependency specialization effectively reduces the share of third-party code in  Java projects.

As future work we will investigate dependency specialization to increase diversity in software supply chains \cite{balliu2023challenges}.
\deptrim currently generates one specialized dependency tree for each project. However, there exists a multitude of possibilities within the realm of partially specialized trees, which we have yet to explore. 
We will venture into the forest of dependency trees to let diversity blossom in Java applications.


\section*{Acknowledgements}
\noindent This work has been partially supported by the Wallenberg Autonomous Systems and Software Program (WASP) funded by the Knut and Alice Wallenberg Foundation, as well as by the Chains project funded by the Swedish Foundation for Strategic Research (SSF).

\IEEEtriggeratref{0}
\balance
\bibliographystyle{ieeetr}
\bibliography{biblio/biblio}

\newpage

\begin{IEEEbiography}[{\includegraphics[width=1in,height=1in,clip,keepaspectratio]{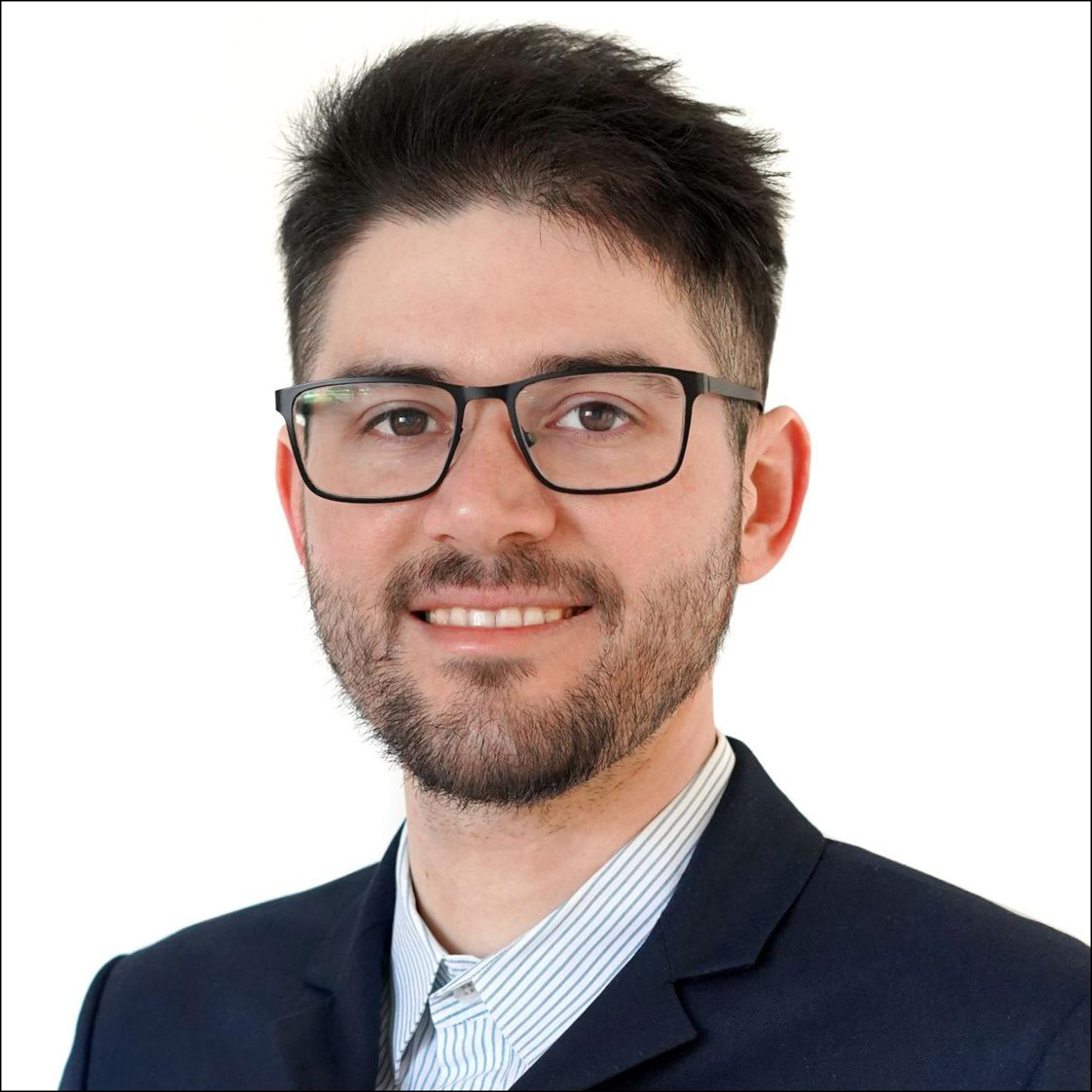}}]%
{C\'esar Soto-Valero} is currently a software engineer working in the financial sector in Stockholm, Sweden. His research interests include code quality, code evolution, and blockchain technologies. Soto-Valero received a Ph.D. in Computer Science from the KTH Royal Institute of Technology in 2023.
\end{IEEEbiography}

\begin{IEEEbiography}[{\includegraphics[width=1in,height=1in,clip,keepaspectratio]{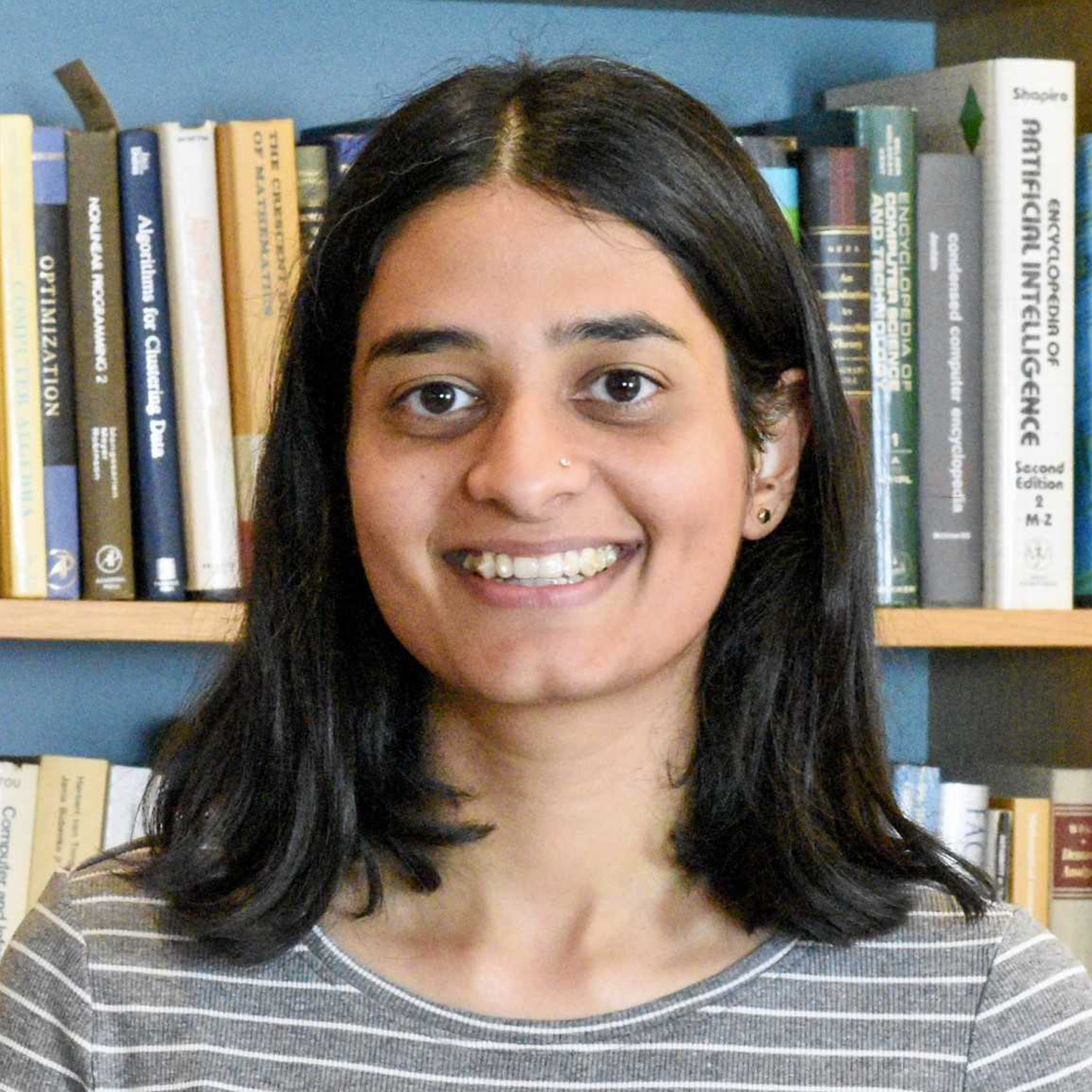}}]%
{Deepika Tiwari} is a Ph.D. student at KTH Royal Institute of Technology, working on software testing. Her research focuses on automatic software test generation, production monitoring, and software humor.
\end{IEEEbiography}

\begin{IEEEbiography}[{\includegraphics[width=1in,height=1in,clip,keepaspectratio]{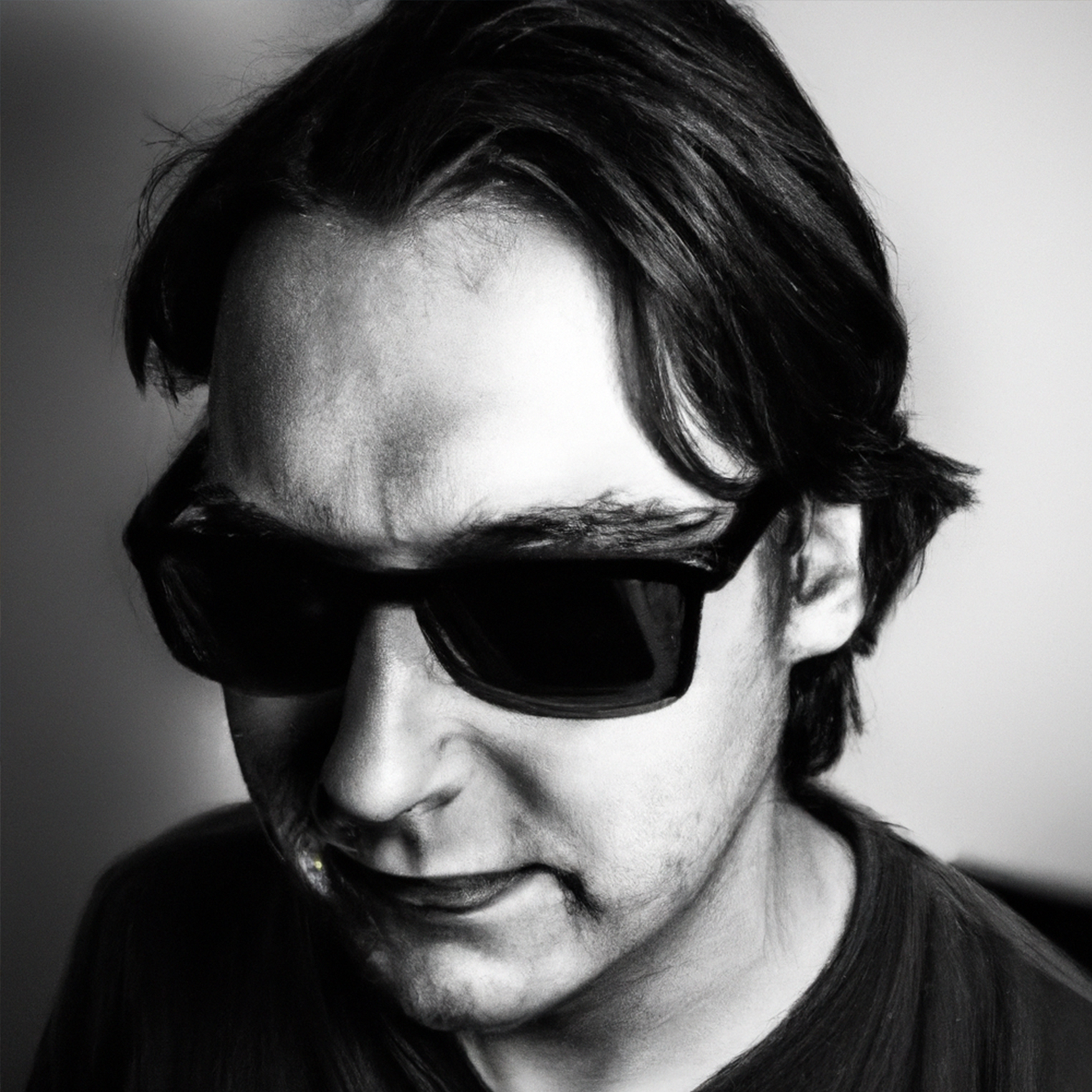}}]%
{Tim Toady} is a Ruby developer and analyst living in Merise, Estonia. His research interests include code analysis for web applications, paint splatters and Easter eggs.
\end{IEEEbiography}

\begin{IEEEbiography}[{\includegraphics[width=1in,height=1in,clip,keepaspectratio]{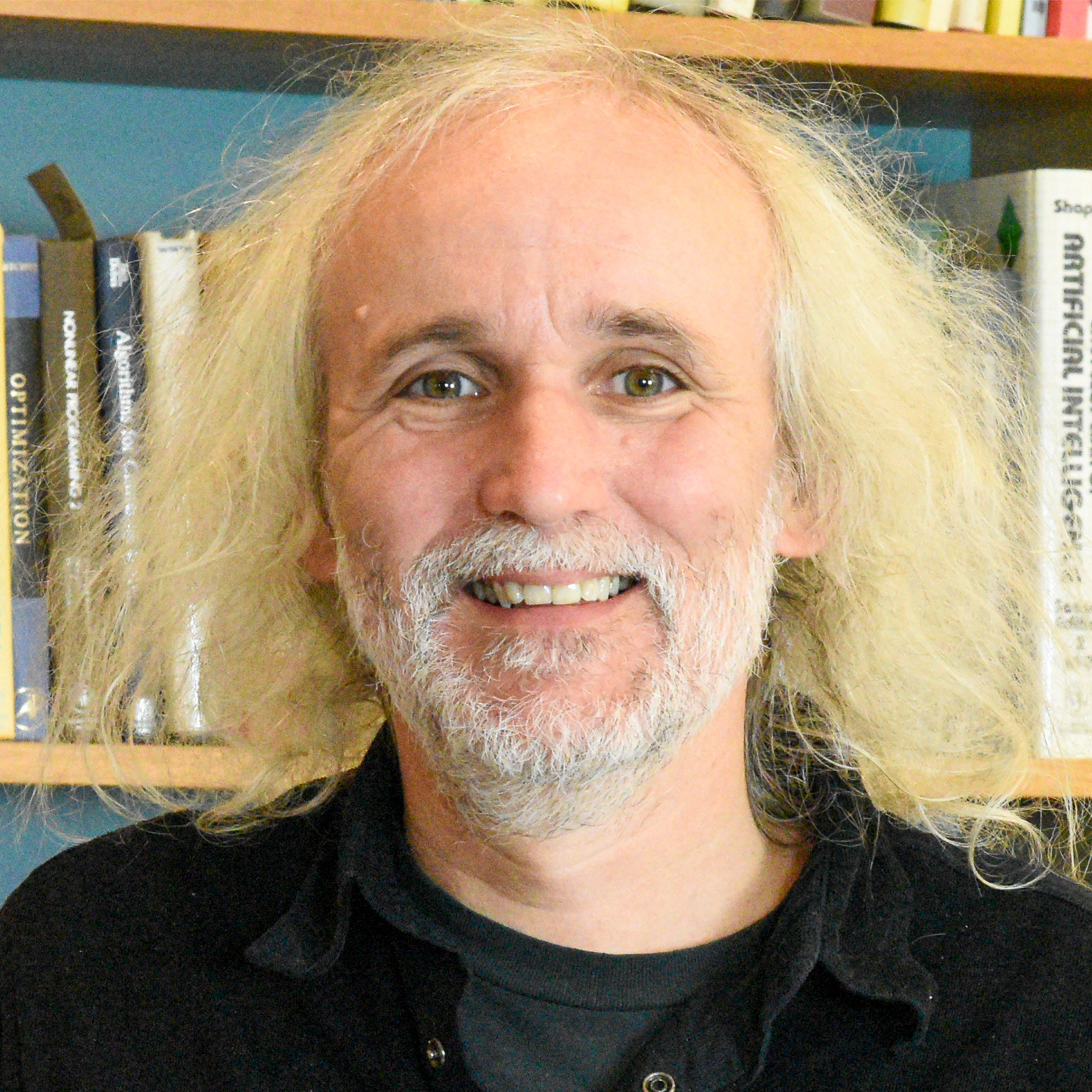}}]%
{Benoit Baudry} is a Professor in Software Technology at the KTH Royal Institute of Technology. His research focuses on automated software engineering, software diversity and software testing. He favors exploring code execution over code on disk and has a genuine interest in the cultural aspects of software development, focusing on humor and art. He received his Ph.D. in 2003 from the University of Rennes, France and was a research scientist at INRIA from 2004 to 2017.
\end{IEEEbiography}

\end{document}